\documentclass[prd,amsmath,
twocolumn,floatfix,amssymb, preprintnumbers, nofootinbib, superscriptaddress]{revtex4}

\usepackage{amssymb, amsmath, amsfonts, bm}
\usepackage{euscript}
\usepackage{float}

\usepackage[colorlinks=true, pdfstartview=FitV, linkcolor=blue, citecolor=blue, urlcolor=blue]{hyperref}

\usepackage{pstricks,pst-node,pst-text,pst-3d}
\usepackage{graphicx}
\usepackage{color}
\usepackage[normalem]{ulem}
\usepackage{relsize} 
\usepackage{dsfont}
\usepackage[version=4]{mhchem}
\usepackage[makeroom]{cancel}
\usepackage{mathtools}
\usepackage{mathrsfs}
\usepackage{booktabs,tabulary}
\usepackage{mciteplus}
\usepackage{siunitx}

\DeclareMathOperator{\im}{Im}

\newcommand*\diff{\mathop{}\!\mathrm{d}}

\newcommand{\be}{\begin{equation}}
\newcommand{\ee}{\end{equation}}
\newcommand{\bml}{\begin{multline}}
\newcommand{\eml}{\end{multline}}
\newcommand{\BM}{\begin{pmatrix}}
\newcommand{\EM}{\end{pmatrix}}

\begin{document}

\title{Nuclear coherent $\pi^0$ photoproduction with charge-exchange and spin-flip rescattering}

\author{V. Tsaran}
\email{vitsaran@uni-mainz.de}
\author{M. Vanderhaeghen}
\address{Institut f\"ur Kernphysik \& PRISMA$^+$  Cluster of Excellence, Johannes Gutenberg Universit\"at,  D-55099 Mainz, Germany}

\date{\today }

\begin{abstract}
In this work, we present an updated model for nuclear $\pi^0$ photoproduction, which incorporates pion second-order rescattering on intermediate excited nuclear states. 
Our approach is based on the distorted wave impulse approximation in momentum space.
The many-body medium effects are incorporated in the complex effective $\Delta$ self-energy, employing the results of the recently developed second-order pion-nucleus scattering potential.
The experimental data for ${}^{12}$C and ${}^{40}$Ca are successfully described without the need to fit the model parameters of the photoproduction amplitude as a result of incorporating the second-order part of the nuclear photoproduction potential, which involves intermediate nucleon spin-flip and charge exchange.
\end{abstract}

\maketitle

\section{Introduction}
\label{sec:intro}
About a decade after the pion was discovered in the late 1940s, coherent photoproduction of $\pi^0$ was first proposed as a valuable tool for studying the distribution of nucleons in nuclei~\cite{leiss1958nuclear, Schrack:1962zz}.
While this physics motivation remains relevant nowadays~\cite{Krusche:2005jx}, the predominant focus has shifted towards extracting the neutron skin of heavy nuclei~\cite{Tarbert:2013jze, Bondy:2015uoa}, which is the difference between the root-mean-square radii of the neutron and proton distributions. 
The neutron skin measurement provides a unique constraint on the symmetry energy of the nuclear Equation Of State, which strongly depends on poorly constrained three-body forces~\cite{Tsang:2012se, Horowitz:2014bja}.
Consequently, an accurate extraction of the neutron skin in neutron-rich nuclei is motivated for the understanding of neutron star structure.
However, Ref.~\cite{Colomer:2022nqi} demonstrated low sensitivity of the $\pi^0$ coherent photoproduction cross section to variations in the neutron skin thickness, which underscores the importance of a very good understanding of the theoretical model for accurately extracting the neutron skin~\cite{Thiel:2019tkm}.

Furthermore, nuclear coherent $\pi^0$ photoproduction is an effective tool for investigating the properties of the pion-nucleus interaction.
For this reason, this process finds relevance in long-baseline neutrino oscillation experiments, e.g., T2K~\cite{T2K:2011qtm} and  Hyper-Kamiokande~\cite{Hyper-Kamiokande:2018ofw}, where quasifree single-pion production is a dominant process within the energy region corresponding to the $\Delta(1232)$-resonance excitation.
Although the coherent $\pi^0$ production in neutrino-nucleus collisions is not a dominant process, it helps to explore the quasifree pion production.
Coherent $\pi^0$ photoproduction may serve as a reliability test for an approach utilized to describe coherent pion production in neutrino-nucleus scattering~\cite{Nakamura:2009iq}.
While the coherent $\pi^0$ neutrino-nucleus production amplitude requires the incorporation of the hadronic axial-vector current component in the elementary amplitude, the treatment of the vector part aligns with that of electro- and photoproduction.

Previously, various theoretical approaches have extensively explored the coherent $\pi^0$ photoproduction on spin-zero nuclei, each with distinct focuses~\cite{Drechsel:1999vh, Peters:1998mb, Abu-Raddad:1998ams}.
As in the case of the pion-nucleus scattering process, we can distinguish two major groups of the existing models.
The first group encompasses the distorted wave impulse approximation (DWIA) models, e.g., Refs.~\cite{Girija:1983qm, Boffi:1986yj, Chumbalov:1987js}, which primarily investigate the final state interaction of the pion propagating within the nuclear medium in a momentum-space representation.
The DWIA in momentum space was first applied in Ref.~\cite{Eramzhian:1983pe} for the $\pi^+$ photoproduction on ${}^{16}$O.
These models dynamically treat the pion-nucleus interaction, taking into account nonlocality and off-shell effects.
Nevertheless, they typically do not consider the medium-induced shifts of the  $\Delta$-resonance's mass and width.
The second group of the $\Delta$-hole models, e.g., Refs.~\cite{Saharia:1980dm, Oset:1981wj, Koch:1982fj, Koerfgen:1994gg, Carrasco:1991we}, extensively probes the in-medium characteristics of the $\Delta$ isobar, mainly focusing on the $\Delta$ and pion dynamics but often neglecting nonresonant contributions. 
However, numerically complex $\Delta$-hole calculations have mainly been limited to light nuclei.
Reference~\cite{Drechsel:1999vh} effectively merged these two approaches by employing the DWIA calculation alongside the Unitary Isobar Model~\cite{drechsel1999unitary} for the elementary photoproduction amplitude. 
The in-medium effects were incorporated in the phenomenological $\Delta$ self-energy, fitted to experimental data for $\pi^0$ photoproduction on ${}^4$He from Ref.~\cite{Rambo:1999jz}.
Subsequently, this model has been successfully applied to describe coherent $\pi^0$ photoproduction data across a diverse range of nuclei~\cite{Bellini:1999nm, Krusche:2002iq, Tarbert:2007whk}.

Utilizing the theoretical framework outlined in Ref.~\cite{Drechsel:1999vh} to analyze data on the coherent $\pi^0$ photoproduction off ${}^{208}$Pb measured with the Crystal Ball at the Mainz Microtron (MAMI), the A2 Collaboration derived the neutron skin thickness of ${}^{208}$Pb to be $0.15 \pm 0.03 \,(\text{stat.})_{-0.03}^{+0.02} \,\text{(sys.)} \, \text{fm}$~\cite{Tarbert:2013jze}.
This outcome strongly contrasts with the value of $0.283 \pm \SI{0.071}{fm}$ obtained from the state-of-the-art parity violation measurement in electron scattering by the PREX-2 Collaboration~\cite{PREX:2021umo}, as well as findings from other methods used previously~\cite{Klos:2007is, Zenihiro:2010zz, Tamii:2011pv}.
The discrepancy between these results could be attributed to several contributions ignored in the model of Ref.~\cite{Drechsel:1999vh}, which was used in the analysis of Ref.~\cite{Tarbert:2013jze}.
Such neglected contributions include intermediate pion-nucleon charge exchange~\cite{Gardestig:2015eca} and nucleon spin-flip.
As demonstrated in Ref.~\cite{Miller:2019btv}, incorporating the final-state charge-exchange effects can lead to an additional 5–6\% increase in the predicted $\pi^0$ photoproduction cross sections, significantly impacting the determination of the neutron skin.

The mechanism of pion rescattering on a nucleon via charge-exchange and/or spin-flip interactions holds significant importance in both nuclear pion photoproduction and scattering processes. 
The charge-exchange yields an essential contribution for describing the $\pi^0$ photoproduction on the deuteron and ${}^3$He~\cite{Argan:1980zz, Laget:1981jq, Wilhelm:1996ed}.
Reference~\cite{Odagawa:1991nj} demonstrated the substantial influence of charge-exchange and spin-flip rescattering effects on the differential cross sections of $\gamma$-induced charged pion production on ${}^{12}$C, ${}^{13}$C, and ${}^{15}$N.
Notably, the $\pi^+$ photoproduction on ${}^{12}$C was shown to be less affected by the rescattering processes, with the spin-flip effect being much more significant than charge exchange.
This distinction from the $\pi^-$ photoproduction on ${}^{13}$C and ${}^{15}$N  underscores the strong sensitivity of spin- and isospin-dependent rescattering processes on the nuclear structure.
The second-order charge exchange part of the $s$-wave pion-nucleus scattering potential initially introduced in Ref.~\cite{Ericson:1966fm} has been proven essential for both pion scattering~\cite{Friedman:2004jh} and pionic atoms~\cite{Friedman:2002sys}.
Furthermore, this contribution was shown to be significant also for the $p$-wave part of the scattering potential \cite{Stricker-thesis}.
Finally, in our recent work for pion-nucleus scattering~\cite{Tsaran:2023qjx}, the inclusion of charge-exchange and spin-flip second-order rescattering was demonstrated to yield sizable corrections in the $\Delta$-resonance energy region.

This paper builds upon our prior research on pion-nucleus scattering.
We aim to describe the nuclear $\pi^0$ photoproduction within the framework established in Ref.~\cite{Tsaran:2023qjx}, utilizing the common effective description of the $\Delta$ self-energy for both processes.
For this purpose, we develop the second-order momentum-space potential for nuclear $\pi^0$ photoproduction.
While the first-order part of the potential, obtained within the framework of the impulse approximation, has a standard form~\cite{Chumbalov:1987js, Drechsel:1999vh}, the second-order part is constructed based on our recently developed second-order pion-nucleus scattering potential~\cite{Tsaran:2023qjx}.
This second-order contribution of both pion scattering and photoproduction potentials describes pion rescattering on intermediate excited nuclear states.
It takes into account the explicit spin and isospin structure of the elementary scattering and photoproduction amplitudes involving intermediate nucleon spin-flip and charge-exchange interactions.
In the present calculation of coherent $\pi^0$ photoproduction on ${}^{12}$C and ${}^{40}$Ca, the in-medium effects are incorporated using the approach of Ref.~\cite{ Drechsel:1999vh}.
However, to determine the numerical value of the effective $\Delta$ self-energy, which modifies the elementary photoproduction amplitude, we directly employ the results obtained from the fit of $\pi^\pm$-${}^{12}$C scattering data performed in Ref.~\cite{Tsaran:2023qjx}, covering the energy range of 80–\SI{180}{MeV} pion laboratory kinetic energy.

The paper is organized as follows:
In Sec.~\ref{sec:multiple-scattering}, we present the main aspects of the multiple-scattering formalism for nuclear pion photoproduction.
Then, in Sec.~\ref{sec:pion-nucl-photo-ampl}, we consider the structure of the elementary amplitude of pion photoproduction and its modification in the nuclear medium.
In Sec.~\ref{sec:Vgamma}, we derive the second-order potential for coherent nuclear pion photoproduction.
Next, in Sec.~\ref{sec:results-photo}, we compare the predictions of our photoproduction model with experimental data for ${}^{12}$C and ${}^{40}$Ca nuclei.
Finally, in Sec.~\ref{sec:conclusion}, we present our conclusions.

\section{multiple-scattering formalism}
\label{sec:multiple-scattering}

The $T$-matrix for nuclear pion photoproduction can be presented in terms of a multiple-scattering series:
\begin{multline}
\hat T_\gamma = \sum_{i=1}^A \hat \tau_i^\gamma + \sum_{i=1}^A \sum_{j \ne i}^A \hat \tau_j \hat G  \hat \tau_i^\gamma  \\
+ \sum_{i=1}^A \sum_{j \ne i}^A \sum_{k \ne j}^A \hat \tau_k \hat G \hat \tau_j \hat G \hat \tau_i^\gamma + \cdots,
\label{T-gamma-series}
\end{multline}
where $A$ is the nucleon number, 
$\hat \tau_i$ and $\hat \tau_i^\gamma$ are transition amplitudes for pion scattering and photoproduction on a single nucleon inside the nucleus, respectively, and $\hat G$ is the Green's function of the noninteracting pion-nucleus system.

Due to the structural similarity between Eq.~(\ref{T-gamma-series}) and the series describing the pion-nucleus scattering, the Kerman-McManus-Thaler (KMT) multiple-scattering approach~\cite{Kerman:1959fr} can be applied to the pion photoproduction process~\cite{Kamalov:1996qf}.
Consequently, Eq.~(\ref{T-gamma-series}) can be subdivided into a system of equations:
\begin{subequations}
\begin{align}
&\hat T_\gamma = \hat U_\gamma + \frac{A-1}A \hat T \hat G \hat P_0 \hat U_\gamma,
\label{Ugamma_series}
\\
&\hat T = \hat U + \frac{A-1}A \hat U \hat G \hat P_0 \hat T,
\label{U_series}
\end{align}
\label{U+Ugamma_series}%
\end{subequations}
where $T$ is the transition matrix for pion-nucleus scattering, while $\hat U_\gamma$ and $\hat U$ are the pion-nuclear photoproduction and scattering potentials, respectively, defined as
\begin{subequations}
\begin{align}
&\hat U_\gamma = 
A \hat \tau_1^\gamma + A(A - 1) \hat \tau_2 \hat G \hat P_\emptyset \hat \tau_1^\gamma 
\notag
\\
& \qquad\qquad\qquad 
+ A(A - 1)^2 \hat \tau_3  \hat G \hat P_\emptyset \hat \tau_2 \hat G \hat P_\emptyset \hat \tau_1^\gamma + \cdots,
\label{Ugamma(tau)-series}
\\
&\hat U = 
A \hat \tau_1 + A(A - 1) \hat \tau_2  \hat G \hat P_\emptyset \hat \tau_1 
\notag
\\
& \qquad\qquad\qquad 
+ A(A - 1)^2 \hat \tau_3 \hat G \hat P_\emptyset \hat \tau_2  \hat G \hat P_\emptyset \hat \tau_1 + \cdots.
\label{U(tau)-series}
\end{align}
\label{U+Ugamma(tau)-series}%
\end{subequations}
In these equations, we have used the fact that the ground state wave function of the target nucleus, $|\Psi_0 \rangle$, is completely antisymmetrized, and each term of the same order in the series gives an equal contribution.
The operators $\hat P_0 = |\Psi_0 \rangle \langle \Psi_0|$ and $\hat P_\emptyset =  \hat{ \mathds{1} } - \hat P_0$ project on the nuclear ground state and all possible excited states, respectively.
The factor $(A - 1)/A$ in the equations above prevents double counting of the rescattering since both $\hat \tau_i$ and $\hat \tau_i^\gamma$ already include pion scattering to all orders on a single nucleon.

The standard KMT equations for pion-nucleus scattering are given by Eqs.~(\ref{U_series}) and~(\ref{U(tau)-series}).
As can be seen,  Eqs.~(\ref{Ugamma_series}) and~(\ref{Ugamma(tau)-series}) describing the pion photoproduction are reduced to the corresponding equations for scattering by removing superscript "$\gamma$".
This convention in the notations will also be valid for all quantities in the following.

\begin{figure}[!t]
\center{\includegraphics[width=0.8\linewidth]{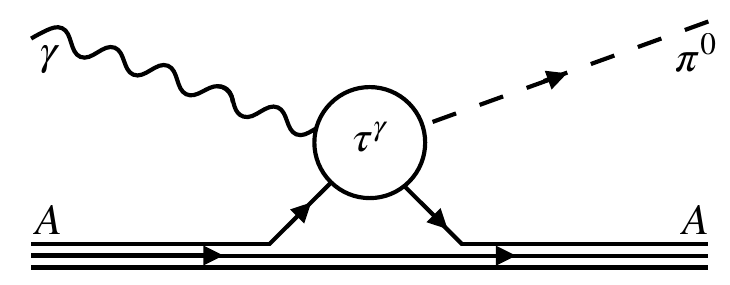}}
\center{\includegraphics[width=0.8\linewidth]{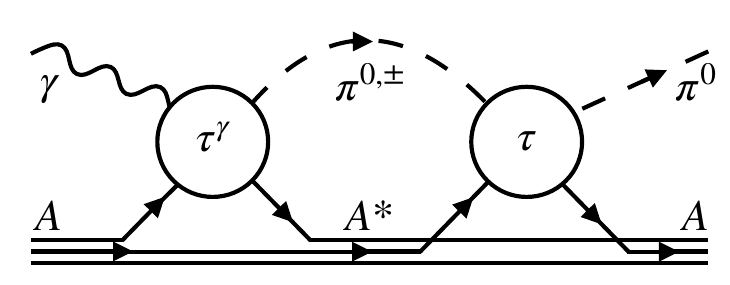}}
\center{\includegraphics[width=0.8\linewidth]{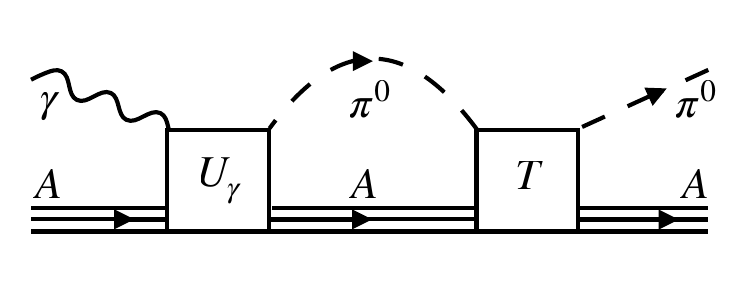}}
\caption{
Diagrammatic representation of the KMT formalism, Eqs.~(\ref{U+Ugamma_series}) and~(\ref{U+Ugamma(tau)-series}), applied to the coherent $\pi^0$ photoproduction on the nucleus.
The upper and middle panels depict the first and second terms on the right-hand side of Eq.~(\ref{Ugamma(tau)-series}), respectively.
The lower panel represents the second term on the right-hand side of Eq.~(\ref{Ugamma_series}).
In the bottom line of the diagrams, $A$ and $A^*$ denote the nucleus in its ground and excited states.
}
\label{fig-Tgamma-diagrams}
\end{figure}

For nuclear coherent $\pi^0$ photoproduction and elastic pion scattering processes, only the nuclear ground state expectation values of the transition amplitudes ($\hat T_\gamma$ and $\hat T$) and the potential operators ($\hat U_\gamma$ and $\hat U$) contribute to Eqs.~(\ref{U+Ugamma_series}). 
The involvement of excited intermediate nuclear states is excluded by the action of the projector $\hat P_0$.
All information about nuclear excitation, including intermediate nucleon spin-flip and charge exchange, is encapsulated in the nuclear pion scattering and photoproduction potential operators, given by Eqs.~(\ref{U+Ugamma(tau)-series}).
Figure~\ref{fig-Tgamma-diagrams} illustrates the diagrams corresponding to the nuclear coherent $\pi^0$ photoproduction. 
The upper panel illustrates the pion photoproduction on a nucleon within the nucleus, described by the first term on the right-hand side of Eq.~(\ref{Ugamma(tau)-series}).
The middle panel corresponds to the second term on the right-hand side of Eq.~(\ref{Ugamma(tau)-series}), representing $\pi^0$ photoproduction on one nucleon, after which the nucleus undergoes a transition to an excited state, followed by propagation and then scattering on a second nucleon.
Finally, the second term on the right-hand side of Eq.~(\ref{Ugamma_series}) represents the rescattering of the outgoing $\pi^0$ off the nucleus, corresponding to the lower panel in Figure~\ref{fig-Tgamma-diagrams}.
The pion elastic scattering process is described by the same set of diagrams with the photon line replaced with the pion one.

As follows from Eq.~(\ref{Ugamma_series}), the coherent $\pi^0$ photoproduction amplitude off nuclei can be presented in momentum space as
\begin{multline}
F_\gamma(\bm k^\prime, \bm k) = V_\gamma(\bm k^\prime, \bm k) \\
- \frac{A - 1}A  \int \frac{\diff \bm k^{\prime\prime}}{2\pi^2} \frac{F(\bm k', \bm k'') V_\gamma(\bm k'', \bm k)}{k_0^2 - {k''}^2 + i \, \varepsilon},
\label{Fgamma-final}
\end{multline}
where $\bm k$  ($\bm k'$) is the incident photon (emitted pion) momentum, and $k_0$ is the pion on-shell momentum. 
The momentum space coherent photoproduction amplitude $F_\gamma$, elastic scattering amplitude $F$, and photoproduction potential $V_\gamma$ are  related to the ground state matrix elements of the corresponding operators via
\begin{subequations}
\begin{align}
&F_\gamma(\bm k', \bm k) = - \frac{\sqrt{\mathscr{M}(k') \mathscr{M}_\gamma(k)}}{2\pi} \langle \pi(\bm k'), \Psi_0 | \hat T_\gamma | \gamma(\bm k), \Psi_0 \rangle,
\\
&V_\gamma(\bm k', \bm k) = - \frac{\sqrt{\mathscr{M}(k') \mathscr{M}_\gamma( k)}}{2\pi} \langle \pi(\bm k'), \Psi_0 | \hat U_\gamma | \gamma(\bm k), \Psi_0 \rangle,
\label{V_gamma-def}
\\
&F(\bm k', \bm k) = - \frac{\sqrt{\mathscr{M}(k') \mathscr{M}(k)}}{2\pi}  \langle \pi(\bm k'), \Psi_0 | \hat T | \pi(\bm k), \Psi_0 \rangle.
\end{align}
\label{FV_gamma-F-def}%
\end{subequations}
The off-shell analog of the relativistic reduced mass of the pion-nucleus system $\mathscr{M}$ in the pion-nucleus center-of-mass (c.m.) frame is defined as
\begin{multline}
\mathscr{M}(k) = \\ \frac{[E + \omega(k) + E_A(k)][\omega(k_0)E_A(k_0) + \omega(k)E_A(k)]}{2 \left(E^2 + (\omega(k) + E_A(k))^2\right)},
\end{multline}
where $\omega(k)$  and $E_A(k)$ are the relativistic energy of the pion and the nucleus, respectively, depending on the c.m. momentum $k$, and $E = \omega(k_0) + E_A(k_0) = k_0^\gamma + E_A(k_0^\gamma)$ is the reaction energy.
For the on-shell photon of momentum $|\bm k| = k_0^\gamma$, $\mathscr{M}_\gamma$ is the simple photon-nuclear relativistic reduced mass:
$\mathscr{M}_\gamma(k) = k_0^\gamma E_A(k_0^\gamma) / E$.

According to Eq.~(\ref{Fgamma-final}), knowledge of both the photoproduction potential and the scattering amplitude is necessary to derive the full photoproduction amplitude.
In this work, we employ the scattering amplitude $F(\bm k', \bm k)$ obtained from Ref.~\cite{Tsaran:2023qjx}, where the second-order pion-nucleus scattering potential incorporating the explicit spin and isospin structure of the elementary scattering amplitude in pion rescattering was fitted to $\pi^\pm$-${}^{12}$C scattering data.
As follows from Eq.~(\ref{Ugamma(tau)-series}), the ground state expectation value of the photoproduction potential operator, $\langle \Psi_0 | \hat U_\gamma | \Psi_0 \rangle$, involves an infinite series of terms. 
All the terms, except the first one, consist of nondiagonal matrix elements of $\hat \tau$ and $\hat \tau_\gamma$ due to the presence of the projector on excited states, $\hat P_\emptyset$.
As in the case of the pion-nucleon scattering process, we assume that the photoproduction potential is approximated by the first two terms corresponding to the two top panels of Fig.~\ref{fig-Tgamma-diagrams}:
\be
\hat U_\gamma \approx
A \hat \tau_1^\gamma + A(A - 1) \hat \tau_2  \hat G \hat P_\emptyset \hat \tau_1^\gamma.
\label{Ugamma-2term}
\ee

Further development of the nuclear photoproduction potential relies on a comprehensive knowledge of the amplitude for the reaction of photoproduction on a single nucleon,  
\be
\gamma(\bm k) + N(\bm p) \longrightarrow \pi(\bm k') + N(\bm p'),
\ee
covering its spin-isospin structure, off-shell behavior, and modification in the nuclear medium, which will be discussed next.

\section{Elementary pion photoproduction amplitude}
\label{sec:pion-nucl-photo-ampl}

In general, $\hat \tau^\gamma$ and $\hat \tau$ are energy-dependent $(A+1)$-body operators, related to the amplitudes in free space as
\be
\hat \tau^\gamma = \hat \tau^\gamma(E) = \hat t^\gamma(W) + \hat \tau(E) \left[ \hat G(E) -  \hat g(W)\right]\hat t^\gamma(W),
\ee
where $W$ is the pion-nucleon reaction energy, $\hat t^\gamma$ is the elementary pion photoproduction amplitude on the free nucleon, and $\hat g(W)$ is the propagator of the free pion-nucleon system.
Here and further, we drop the index "i" when there is no need to distinguish nucleons.
As in the case of the pion scattering process, we apply the \textit{impulse approximation}, assuming  $\hat \tau_i^\gamma(E) \approx \hat t^\gamma(W)$.
There are various approaches, each motivated by different considerations, to determine the optimal value for $W$~\cite{Gmitro:1985kj, Chumbalov:1987gd, mach1983galileo, Gurvitz:1986zza}.
For consistency with our previous study of pion scattering~\cite{Tsaran:2023qjx}, we adopt the so-called \textit{two-body choice for the reaction energy}:
\be
W = \sqrt{\left(k + E_N(p) \right)^2 - (\bm k + \bm p)^2},
\label{W-def}
\ee
where $E_N(p)$ is the energy of the struck nucleon.
The reaction energy, Eq.~(\ref{W-def}), is evaluated for the on-shell momenta $|\bm k| = k_0^\gamma$ and $|\bm k'| = k_0$.

The momentum space $T$-matrix for pion photoproduction on free nucleon is related to the free nucleon photoproduction amplitude $f^\gamma$ in the photon-nucleon c.m. frame as 
\begin{multline}
\langle \pi(\bm k^\prime), N(\bm p^\prime) | \hat t^{\gamma} | \gamma(\bm k), N(\bm p) \rangle = 
- (2\pi)^3
\delta(\bm k^\prime + \bm p^\prime - \bm k - \bm p) 
\\
\times
\frac{2\pi}{\sqrt{\mu(\bm k', \bm p')\mu_\gamma(\bm k, \bm p)}} \hat f^{\gamma}(\bm k'_\text{2cm}, \bm k_\text{2cm}^{\vphantom{\prime}}) ,
\label{<tgamma>-fgamma()}
\end{multline}
where $\mu(\bm k, \bm p) = \omega(k) E_N(p) / W(\bm k, \bm p)$, $\mu_\gamma(\bm k, \bm p) = k E_N(p) / W(\bm k, \bm p)$, and $W(\bm k, \bm p)$ is the invariant mass of the system with momenta $\bm k$ and $\bm p$.
The subscript "2cm" indicates the photon-nucleon c.m. frame.

The momenta are boosted to the 2cm frame by the Lorentz transformation
\begin{subequations} 
\begin{align}
&\bm k_\text{2cm}  = \bm k
+ \frac{(\bm k + \bm p)}{W(\bm k, \bm p)} 
\notag \\ 
& \qquad\qquad\qquad \times
\left(\frac{(\bm k + \bm p) \cdot \bm k}{W(\bm k, \bm p) + k + E_N(p)} - k \right), \\
&\bm k_\text{2cm}'  = \bm k' 
+ \frac{(\bm k' + \bm p')}{W(\bm k', \bm p')} 
\notag \\ 
& \qquad  \times
\left(\frac{(\bm k' + \bm p') \cdot \bm k'}{W(\bm k', \bm p') + \omega(k') + E_N(p')} - \omega(k') \right).
\end{align}
\label{Lorentz-transform-gamma}%
\end{subequations}
While the Lorentz transformation is originally derived for on-mass-shell objects, we assume Eqs.~(\ref{Lorentz-transform-gamma}) are also applicable to particles off their mass shell~\cite{Ernst:1980my, Heller:1976wd}. 

The dependence of the amplitudes on the struck nucleon momenta $\bm p$ and $\bm p'$ (the so-called nucleon Fermi motion) significantly complicates the calculation.
In the general case, the procedure implies the integration over nucleon momenta to obtain the matrix element in Eq.~(\ref{V_gamma-def}) and requires knowledge of the model-dependent one-body density matrix~\cite{Chumbalov:1987gd}.
A common approach to simplify this problem is evaluating the elementary amplitudes at the effective nucleon momenta
\be
\bm p =  \frac{\bm k' - \bm k}{2} - \frac{\bm k^\prime + \bm k}{2A}
\quad\text{and}\quad
\bm p^\prime = - \frac{\bm k' - \bm k}{2} - \frac{\bm k^\prime + \bm k}{2A}.
\label{p_eff-def}
\ee
This approximation provides numerical results for cross-sections that closely align with the exact values~\cite{Tiator:1980jw} while conserving energy and momenta.  
Originally introduced for pion scattering~\cite{kowalski1963elastic, Landau:1975zb}, this approach has been proven successful in subsequent applications to photoproduction~\cite{Tiator:1984hv, Chumbalov:1987js, Eramzhian:1990vu}.

\subsection{Decomposition of the pion photoproduction amplitude}

The spin structure of the pion photoproduction amplitude in the photon-nucleon c.m. frame is conventionally expressed in terms of the Chew-Goldberger-Low-Nambu (CGLN) amplitudes~\cite{Chew:1957zz}.
For the production process by a real photon, the decomposition has the form
\begin{multline}
\hat f^\gamma(\bm k', \bm k) = 
i\, \hat{\boldsymbol \sigma} \cdot \boldsymbol \epsilon^\lambda \hat F_1(\bm k', \bm k) \\
+  \frac{(\hat{\boldsymbol \sigma} \cdot \bm k' ) \hat{\boldsymbol \sigma} \cdot [\bm k \times \boldsymbol \epsilon^\lambda]}{k' k} \hat F_2(\bm k', \bm k)  \\
+ 
i \frac{(\hat{\boldsymbol \sigma} \cdot \bm k) (\bm k' \cdot \boldsymbol \epsilon^\lambda)}{k' k} \hat F_3(\bm k', \bm k)
\\
+ i \frac{(\hat{\boldsymbol \sigma} \cdot \bm k')(\bm k' \cdot \boldsymbol \epsilon^\lambda)}{k'^2} \hat F_4(\bm k', \bm k),
\label{CGLN-def}
\end{multline}

where $\boldsymbol \epsilon^\lambda$ is the photon polarization vector, and $\hat{\boldsymbol \sigma}$ is is the nucleon Pauli spin operator.

For circularly polarized photons with the helicity $\lambda = \pm1$ moving in the direction $\bm k_\text{2cm}/k_\text{2cm} = (\sin\theta \cos\phi , \sin\theta \sin\phi , \cos\theta)$, the polarization vector is given by
\begin{multline}
\bm \epsilon^\lambda = \frac{e^{i \lambda \phi}}{\sqrt{2}}  (-\lambda \cos\theta \cos\phi + i \sin\phi, \\
- \lambda \cos\theta \sin\phi - i \cos\phi, \ \lambda \sin\theta),
\end{multline}
satisfying $\bm \epsilon^\lambda \cdot \bm k_\text{2cm} = 0$.

For the on-shell process with a given isospin configuration, the CGLN amplitudes are independent complex functions of the reaction energy and the pion scattering angle. 
The amplitudes $\hat F_i$ can be further expanded into a sum of partial-wave contributions from channels with different pion-nucleon final states with angular momentum $l$.
There are four distinct transition types possible for a real photon, classified based on the parity $P$ of the photon and the final-state total angular momentum $J = |l \pm 1/2|$ of the pion-nucleon system.
For the transverse photon with the total orbital angular momentum $L$, the states can either be electric, $\hat E_{l\pm}$ with $P= (-1)^L$, or magnetic, $\hat M_{l\pm}$ with $P =(-1)^{L+1}$.
The explicit form of the multipole expansion for $\hat F_i$ is given in Ref.~\cite{Chew:1957zz}.

In the computation of the nuclear photoproduction amplitude, as given by Eq.~(\ref{Fgamma-final}), it is necessary to specify the behavior of the multipoles in the case of the off-shell pion-nucleon system.
To maintain consistency with our calculation of the pion–nucleon scattering amplitude in Ref.~\cite{Tsaran:2023qjx}, we employ the following separable form for the pion with the off-shell momentum $\bm k'$:
\be
\hat A_{l\pm}(k',  k_0^\gamma) = \hat A_{l\pm}(k_0, k_0^\gamma)   
\left( \frac{k'}{k_0} \right)^l
{v}(k'), 
\label{A-off-shell}
\ee
where $\hat A_{l\pm}$ encompasses both $\hat M_{l\pm}$ and $\hat E_{l\pm}$. 
The off-shell vertex factor is
\be
{v}(k) = \frac{\Lambda^2 - m_\pi^2}{\Lambda^2 - (\omega^2(k_0) - k^2)},
\label{off-vertex-fact}
\ee
with $\Lambda = 1.25 \text{ GeV}$.
For the on-shell process, $\hat E_{l\pm} = \hat E_{l\pm}(k_0, k_0^\gamma) $ and $\hat M_{l\pm} = \hat M_{l\pm}(k_0, k_0^\gamma)$ depend solely on the reaction energy.

The amplitudes $\hat F_i$, $\hat E_{l\pm}$, and $\hat M_{l\pm}$ are operators in nucleon isospin space.
If isospin conservation in the hadronic system is assumed, the isospin structure of any photoproduction amplitude  involving a pion with a Cartesian isospin index $b$ is given by~\cite{Ericson:1988gk}
\be
\hat A = \delta_{b 3} A^+ + \frac12 [\hat{\pmb \tau}_b,  \hat{\pmb \tau}_3] A^- + \hat{\pmb \tau}_b A^0,
\label{photo-isospin-struct}
\ee
where $\hat{\pmb \tau}_b$ represents the nucleon isospin matrices.
To facilitate the analysis of resonances in the pion-nucleon system with respect to its total isospin $T$, it is convenient to introduce the following combinations of the isovector photon amplitudes:
\begin{subequations}
\begin{align}
&A_{l\pm}^{1/2} = A_{l\pm}^+ + 2 A_{l\pm}^-, \\
&A_{l\pm}^{3/2} = A_{l\pm}^+ - A_{l\pm}^-,
\end{align}
\label{M-isospin-as-pm}%
\end{subequations}
which correspond to $T = 1/2$ and $3/2$, respectively.
The isoscalar amplitudes $A^0$ correspond to isospin-$1/2$ states.

The magnetic dipole amplitude $M_{1+}^{3/2}$ holds significant importance as it represents the amplitude for the photoexcitation of the $\Delta(1232)$ resonance~ \cite{drechsel2007unitary, Workman:2012jf}.
By keeping contributions solely from the $s$- and $p$-wave pion-nucleon system and explicitly separating the $\Delta(1232)$ resonance multipole $M_{1+}^{3/2}$, the half-off-shell CGLN amplitudes are expressed as 
\begin{subequations}
\begin{align}
&F_1^+(\bm k', \bm k) \approx E_{0+}^+ {v}(k') 
\notag\\ 
& \qquad
+ \left(2 M_{1+}^{3/2} + M_{1+}^{1/2} + 3 E_{1+}^{+} \right) \frac{\bm k' \cdot \bm k}{k_0 k_0^\gamma} {v}(k'),\\
&F_1^-(\bm k', \bm k) \approx E_{0+}^- {v}(k') 
\notag\\ 
& \qquad
+ \left(-M_{1+}^{3/2} + M_{1+}^{1/2} + 3 E_{1+}^{-} \right) \frac{\bm k' \cdot \bm k}{k_0 k_0^\gamma} {v}(k'),\\
&F_2^+(\bm k', \bm k) \approx 
\left( \frac43 M_{1+}^{3/2} + \frac23 M_{1+}^{1/2} + M_{1-}^{+} \right) \frac{k'}{k_0} {v}(k'),\\
&F_2^-(\bm k', \bm k) \approx
\left( -\frac23 M_{1+}^{3/2} + \frac23 M_{1+}^{1/2} + M_{1-}^{-} \right) \frac{k'}{k_0} {v}(k'),\\
&F_3^+(\bm k', \bm k) \approx
\left( -2 M_{1+}^{3/2} -  M_{1+}^{1/2} + 3 E_{1+}^{+} \right) \frac{k'}{k_0} v(k'),\\
&F_3^-(\bm k', \bm k) \approx 
\left( M_{1+}^{3/2} -  M_{1+}^{1/2} + 3 E_{1+}^{-} \right) \frac{k'}{k_0} {v}(k'),\\
&F_4^\pm(\bm k', \bm k) \approx 0.
\end{align}
\end{subequations}
A similar expansion is applicable to $F^0_i$, which, however, does not contribute to nuclear $\pi^0$ photoproduction on isospin-zero nuclei. 
In our computation of the first-order part of the photoproduction potential, we also include the $d$-wave contribution in a similar manner, introducing a correction on the order of a few percent at higher energies.

For all multipole amplitudes except $M_{1+}^{3/2}$, we use their free-space values taken from the unitary isobar model MAID2007~\cite{drechsel2007unitary} based on the dynamical Dubna-Mainz-Taipei model~\cite{Kamalov:2000en}.
Note that $E_{0+}^-$ is the dominant nonresonant multipole. 
However, $E_{0+}^-$ does not contribute to the first-order potential for spin-zero nuclei. 
It only appears in the second-order correction (as shown in Sec.~\ref{sec:Vgamma}), and for this reason, its in-medium modification has a negligible impact on observables.

Various $\Delta$-hole model calculations for pion scattering~\cite{Hirata:1978wp, Horikawa:1980cv, Oset:1987re, Freedman:1982yp} and photoproduction~\cite{Koch:1979zz, Saharia:1980dm, Laktineh:1992kp} indicate significant differences between the properties of the $\Delta$ isobar within the nucleus and in free space. 
The decay width of the $\Delta$ is notably affected by numerous inelastic channels opened in the nuclear medium, such as pion absorption on a few nucleons~\cite{Oset:1986yi}.
Correspondingly, the resonant multipole $M_{1+}^{3/2}$ undergoes significant modifications in the case of the pion photoproduction on a bound nucleon, as will be described next.
{
The modification of the much smaller resonant electric quadrupole amplitude $E_{1+}^{3/2}$ is considered in Appendix~\ref{sec:E1p-in-medium}.
}

\subsection{In-medium modification of \texorpdfstring{$M_{1+}^{3/2}$}{MM} 
}
\label{sec:M1p3-in-medium}

The magnetic dipole amplitude $M_{1+}^{3/2}$ consists of a $\Delta$-resonant term and a nonresonant background contribution~\cite{Tanabe:1985rz}:
\be
M_{1+}^{3/2} = M_{1+}^{3/2(B)} + M_{1+}^{3/2(\Delta)}. 
\label{M1p3-B-R-separation}
\ee
In the present calculation, we use the unitary isobar model MAID98~\cite{drechsel1999unitary}, according to which the resonant part of the $M_{1+}^{3/2}$ amplitude can be written as
\begin{multline}
M_{1+}^{3/2(\Delta)} \\
= \sqrt{\frac32} \bar M_{3/2}  \frac{f_{\gamma N \Delta}(W)  f_{\pi N \Delta}(W) \Gamma_\Delta(W) /2 }{m_\Delta - W - i\, \Gamma_\Delta(W) /2}  e^{i \phi(W)},
\label{M1p3-MAID98}
\end{multline}
with {$\bar M_{3/2} = \SI{0.323}{GeV^{-1/2}}$,} the $\Delta$-resonance mass $m_\Delta$ and decay width $\Gamma_\Delta(W)$.
The energy dependence of $\Gamma_\Delta(W)$ in MAID is parametrized as
\be
\Gamma_\Delta(W) = \Gamma_R \frac{k_0^3}{k_R^3} 
\frac{m_\Delta}{W} \frac{2m_\Delta}{W + m_\Delta} 
{
\left(
\frac{X^2 + k^2_R}{X^2 + k^2_0}
\right)^r
}
.
\label{Gamma_Delta}
\ee
The vertex factors are given by
\begin{subequations}
\begin{align}
&f_{\gamma N \Delta}(W) = \left( \frac{k_0^\gamma}{k_R^\gamma} \right)^{n} \left( \frac{X^2 + k_R^{\gamma2}}{X^2 + k_0^{\gamma2}} \right), 
\label{MAID-f_gammaN}
\\
&f_{\pi N \Delta}(W) = \left( \frac1{3\pi} \frac{k_0^\gamma}{k_0} \frac{2 m_N}{W + m_\Delta} \frac1{\Gamma_\Delta(W)}  
\right)^{1/2},
\label{MAID-f_piN}
\end{align}
\label{MAID-vertex-factors}%
\end{subequations}
where $\Gamma_R$ and $k_R^\gamma$  ($k_R$) are the $\Delta$ width and photon (pion) momentum at the resonance position, $W = m_\Delta$. 
In Eqs.~(\ref{MAID-vertex-factors}), $X$ is the damping parameter, $m_N$ is the nucleon mass, {and $n=2$ for $M_{1+}^{3/2}$}. 
The numerical values of the model parameters are listed in the first line of Table~\ref{tabl:MAID-RDIM-pars}.
\begin{table}[htb!]
\caption{ Parameters of MAID98~\cite{drechsel1999unitary} and R$\Delta$M~\cite{Oset:1981ih} 
}
\begin{tabular}{ccccc}
\hline\hline
 &   $m_\Delta$ [MeV] & $\Gamma_R$ [MeV] & \quad {r} \quad & $X$ [MeV] \\
\hline
MAID98 & \quad  1235  & \quad 130 &  \quad {1} \quad  &  \quad 500 \\
R$\Delta$M  & \quad 1232  &  \quad 115 & \quad {0} \quad & \quad -- \\
\hline
\end{tabular}
\label{tabl:MAID-RDIM-pars}
\end{table}

The phenomenological phase $\phi(W)$ is introduced in Eq.~(\ref{M1p3-MAID98}) to fulfill the Fermi-Watson final state theorem~\cite{Watson:1954uc}, which results from the unitarity of the $S$-matrix. 
Following Olson's unitarization procedure~\cite{Olsson:1974sw}, the phase of $M_{1+}^{3/2}$ multipole is adjusted to the pion-nucleon scattering phase shift in the spin-isospin-3/2  $p$-wave channel $P_{33}$, providing the unitary phase
\be
\phi(W) = (22.13 \, x - 3.769 \, x^2 + 0.184 \, x^3) \ [\deg],
\ee
with $x = (W - m_\pi - m_N) / \SI{100}{MeV}$.

\begin{figure*}[!t]
\center{\includegraphics[width=0.8\linewidth]{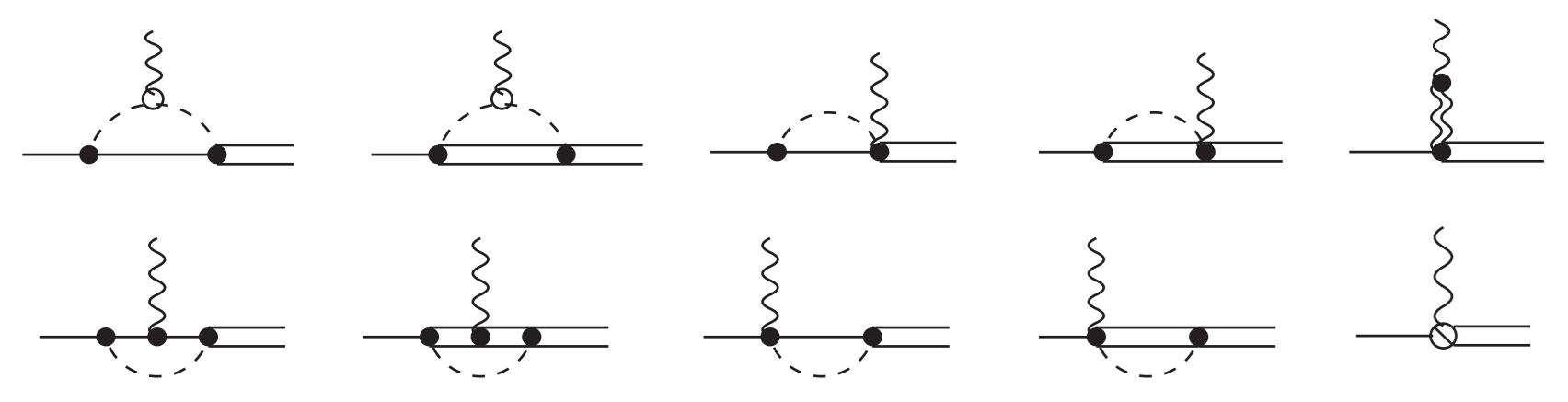}}
\caption{
The NLO $\gamma N \Delta$ vertex corrections in baryon $\chi$EFT in the $\delta$-expansion counting scheme~\cite{Pascalutsa:2006up}.
The solid lines correspond to nucleon propagators; the double lines represent $\Delta$ propagators; the wiggly double line corresponds to the vector–meson propagator.
The latter process appears as the effective low-energy constant associated with the radius of the $g_M$ form factor at finite photon virtuality.
The closed and open circles denote the vertices from the first- and second-order Lagrangian, respectively,
the sliced vertex stands for the electric $g_E$ and Coulomb $g_C$ quadrupole couplings.
}
\label{fig:chiPT-vertex-photo}
\end{figure*}

The separation in Eq.~(\ref{M1p3-B-R-separation}) is not unique and varies, for example, between dynamical models~\cite{Kamalov:2000en, Sato:1996gk, Pascalutsa:2004pk} as well as phenomenological multipole solutions~\cite{drechsel1999unitary, drechsel2007unitary, Workman:2012jf}.
Specifically addressing the $M_{1+}^{3/2}$ multipole and real photon scenarios, the primary distinction between MAID98~\cite{drechsel1999unitary} employed in this work (as well as in Ref.~\cite{Drechsel:1999vh}) and the later unitary isobar model MAID2007~\cite{drechsel2007unitary} is their approach to handling the background.
The background contribution $M_{1+}^{3/2(B)}$ in MAID98, consisting of Born terms and $t$-channel vector-meson contributions, is a real monotonic function.
Within the framework of MAID2007, the background undergoes unitarization through the $K$-matrix approach, transforming into a complex function that has zero value at the resonance position.
Consequently, this difference leads to distinct unitarization phases $\phi(W)$.

The chiral effective field theory ($\chi$EFT)~\cite{Bernard:1995dp} results provide a vigorous framework to understand the role played by the unitary phase.
As was shown in Ref.~\cite{Pascalutsa:2006up} for the pion photoproduction in the $\Delta$-resonance energy region, the Fermi–Watson theorem is fulfilled exactly in the complete next-to-leading-order (NLO) calculation using the $\delta$-expansion power counting scheme~\cite{Pascalutsa:2002pi}.
In this approach, the bare resonant part, which defines the LO result, itself satisfies the Fermi–Watson theorem.
Going to NLO, the background contribution is purely real.
Consequently, their naive summation would result in unitarity breaking.
The restoration of unitarity occurs upon including the $\gamma N \Delta$ vertex corrections illustrated in Fig.~\ref{fig:chiPT-vertex-photo} in the calculation.

Comparing the $\chi$EFT results with the framework of MAID98 reveals that the multiplicative factor $\exp[i\phi(W)]$ in Eq.~(\ref{M1p3-MAID98}) effectively incorporates the contribution of the additive vertex corrections~\cite{Bernstein:1993rv}.
The difference between the approaches of MAID98 and MAID2007 can be related to the vertex correction graphs in Fig.~\ref{fig:chiPT-vertex-photo}.
Depending on whether vertex corrections are associated with the resonant or background contributions, a real or unitary (complex) background term is derived, as implemented in the MAID98 and MAID2007 approaches, respectively.

For the nuclear applications considered in this work, we want to allow for a medium modification of the $\Delta$-resonance. 
{
For this purpose, we adopt the approach of MAID98, where all vertex corrections are absorbed in the resonant part of the $M_{1+}^{3/2}$ multipole, offering a more straightforward process of implementing in-medium correction. Notably, various theoretical frameworks for pion photoproduction on the nucleon have been developed in the last decades, categorized into several major groups: dispersive~\cite{Hanstein:1997tp,  Aznauryan:2002gd}, effective Lagrangian~\cite{Davidson:1991xz, Fernandez-Ramirez:2005nuq, Mariano:2007zza}, dynamical models~\cite{Nozawa:1989gy, Sato:1996gk, Kamalov:2000en, Pascalutsa:2004pk, Caia:2004pm}, chiral effective field theory~\cite{Pascalutsa:2005vq}, and further references therein.  
However,} employing MAID98 enables a direct comparison between the predictions of our model and the results of Ref.~\cite{Drechsel:1999vh}, avoiding discrepancies due to dependence on the model for the photoproduction of free pions.

Our method of introducing the medium effects is then similar to the approach of Ref.~\cite{Drechsel:1999vh}. 
The nonresonant background $M_{1+}^{3/2(B)}$ is assumed to remain unchanged.
The Breit–Wigner denominator and unitary phase in Eq.~(\ref{M1p3-MAID98}) are combined into an effective propagator, which can be presented as
\be
\bar G(W) =  G(W) e^{i\phi} = \frac1{\overline m_\Delta(W) - W - i \, \overline \Gamma_\Delta(W)/2},
\label{bar-G-def}
\ee
with
\begin{subequations}
\begin{align}
&\overline m_\Delta(W) = m_\Delta \cos\phi + 2 W \sin^2 \frac{\phi}2  - \frac12 \Gamma_\Delta(W) \sin\phi, \\
&\overline \Gamma_\Delta(W) =  \Gamma_\Delta(W) \cos\phi - 2(W - m_\Delta) \sin\phi.
\end{align}
\label{barm-barGamma-def}%
\end{subequations}

The effective mass and width of the resonance, Eqs.~(\ref{barm-barGamma-def}), are shifted in the nuclear medium by many-body effects. 
Correspondingly, for the photoproduction off a bound nucleon, the effective propagator, Eq.~(\ref{bar-G-def}), in Eq.~(\ref{M1p3-MAID98}) is replaced by
\be
\bar G(\Sigma_\Delta, W) = \frac1{\overline m_\Delta(W) - W - i \, \overline \Gamma_\Delta(W)/2 + \overline \Sigma_\Delta},
\label{bar-G-modified}
\ee
where $\overline \Sigma_\Delta$ is the effective $\Delta$ self-energy.

In the following subsection, we determine the $\Delta$ self-energy $\overline \Sigma_\Delta$ consistent with our model for pion-nucleus scattering~\cite{Tsaran:2023qjx}, which was fitted to the experimental data.

\subsection{The effective \texorpdfstring{$\Delta$}{M} self-energy}
\label{sec:Delta-self-energy}

Despite the nonlocal nature of the $\Delta$ self-energy $\Sigma_\Delta$ within a finite nucleus~\cite{Oset:1979bi}, an assumption of locality was successfully applied to both scattering~\cite{Hirata:1978wp, Horikawa:1980cv, Oset:1987re} and photoproduction~\cite{Carrasco:1991we} processes within the $\Delta$-hole model framework.
The analysis of low-energy pion-nucleus scattering and pionic atom data, conducted in Ref.~\cite{Seki:1983sh} across a wide range of nuclei, revealed the existence of a nuclear effective density, which determines the parameters of the pion-nucleus potential. 
The evaluation of the average value of $r$ over the product of the nuclear density and the initial and final pion distorted waves reveals that the pion-nucleus scattering predominantly occurs in the peripheral region.
In particular, for the pion-${}^{12}$C elastic scattering at the pion kinetic energy of $\SI{100}{MeV}$, the average value of $r$ is $\SI{2.52}{fm}$ ~\cite{Johnson:1992zw}.
As demonstrated in Ref.~\cite{Oset:1986yi}, the strong two- and three-body pion absorption mechanisms result in nearly complete absorption of pion flux near the nuclear surface, even at the pion laboratory energy of $\SI{165}{MeV}$. 
Only at high pion energies, about $\SI{240}{MeV}$, there is a non-negligible probability that the pions will pass through the nucleus without being absorbed.
Due to the strong absorption processes, the outgoing pions primarily interact with a narrow band of the nuclear surface around the effective nuclear density.
Consequently, it is reasonable to presume consistency in the $\Delta$ self-energy between scattering and photoproduction processes, as only pions produced from the nuclear surface area will likely remain unabsorbed.

For the scattering process in Ref.~\cite{Tsaran:2023qjx}, we adopt the \textit{relativistic $\Delta$-isobar model} (R$\Delta$M)~\cite{Oset:1981ih}, which successfully reproduces the $p$-wave pion-nucleon phase shifts at low and intermediate energies, especially in the resonant $P_{33}$ channel.
The model is based on the $K$-matrix formalism, in which the scattering amplitudes are expressed through
\be
f^l_{2T \, 2J} = \frac{K^l_{2T \, 2J}}{1- i \, k_0 K^l_{2T \, 2J}}.
\label{f(K)}
\ee
We explicitly separate the $s$-channel $\Delta$ contribution to the $p$-wave spin-isospin-$\frac32$ $K$-matrix from the non-resonant background:
\be
K^1_{33} = K^{1 (B)}_{33} + K^{1 (\Delta)}_{33},
\ee
where the background contribution is
\be
K^{1 (B)}_{33} = \frac13 \frac{k_0^2}{m_{\pi}^2} \frac{m_N}{W} 
\frac{f_N^2}{4\pi} \frac{8 m_N}{m_N^2 - \bar u}
\label{K-background}
,
\ee
with $f_N^2/4\pi = 0.079$ and the approximate $u$-channel Mandelstam variable $\bar u = m_N^2 + m_\pi^2 - 2 \omega(k_0) E_N(k_0)$.
In Eq.~(\ref{K-background}), the $u$-channel contributions from the Roper resonance $N^*(1440)$ and $\Delta$ are omitted for simplicity.
The resonant part of the $K$-matrix in the nuclear medium is
\be
K^{1 (\Delta)}_{33} (\Sigma_\Delta) =
\frac1{k_0} \frac{ \Gamma_\Delta(W)/2}{m_\Delta + \Sigma_\Delta - W},
\label{K-in-medium}
\ee
with the effective $\Delta$ self-energy $\Sigma_\Delta$.
Note that R$\Delta$M employs the resonance mass and width slightly distinct from the parameters of MAID98, as listed in the second line of Table~\ref{tabl:MAID-RDIM-pars}.
In our approach, $\Sigma_\Delta$ serves as a complex model parameter independent of energy and momentum.
The fit to $\pi^\pm$-${}^{12}$C scattering data within the energy range of 80–\SI{180}{MeV} pion laboratory kinetic energy yields the numerical value $\Sigma_\Delta = 12.9 \pm 1.3 - i \, (33.2 \pm 0.8) \, \text{MeV}$. 
Employing this value of $\Sigma_\Delta$ leads to reasonable predictions for pion scattering on ${}^{16}$O, ${}^{28}$Si, and ${}^{40}$Ca nuclei.

\begin{figure}[!t]
\center{\includegraphics[width=0.99\linewidth]{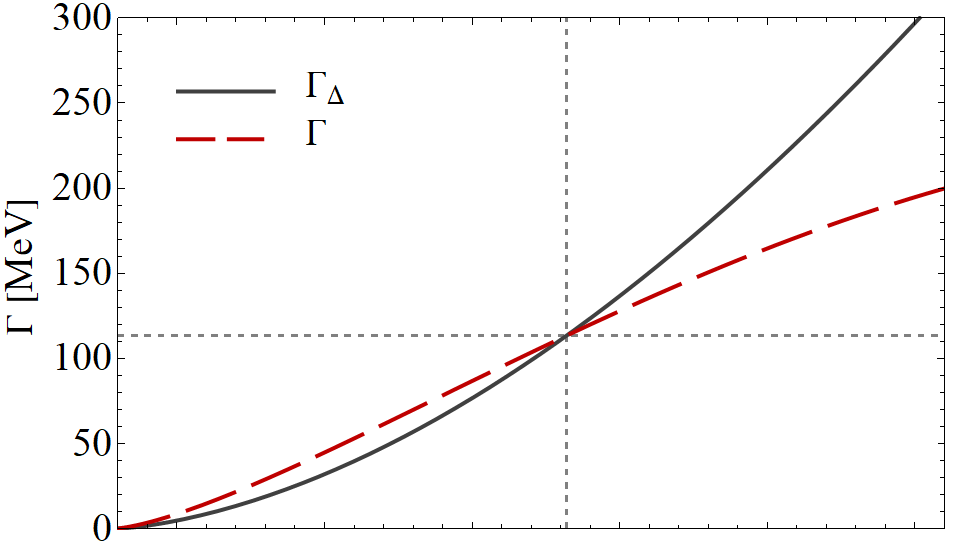}
\includegraphics[width=0.99\linewidth]{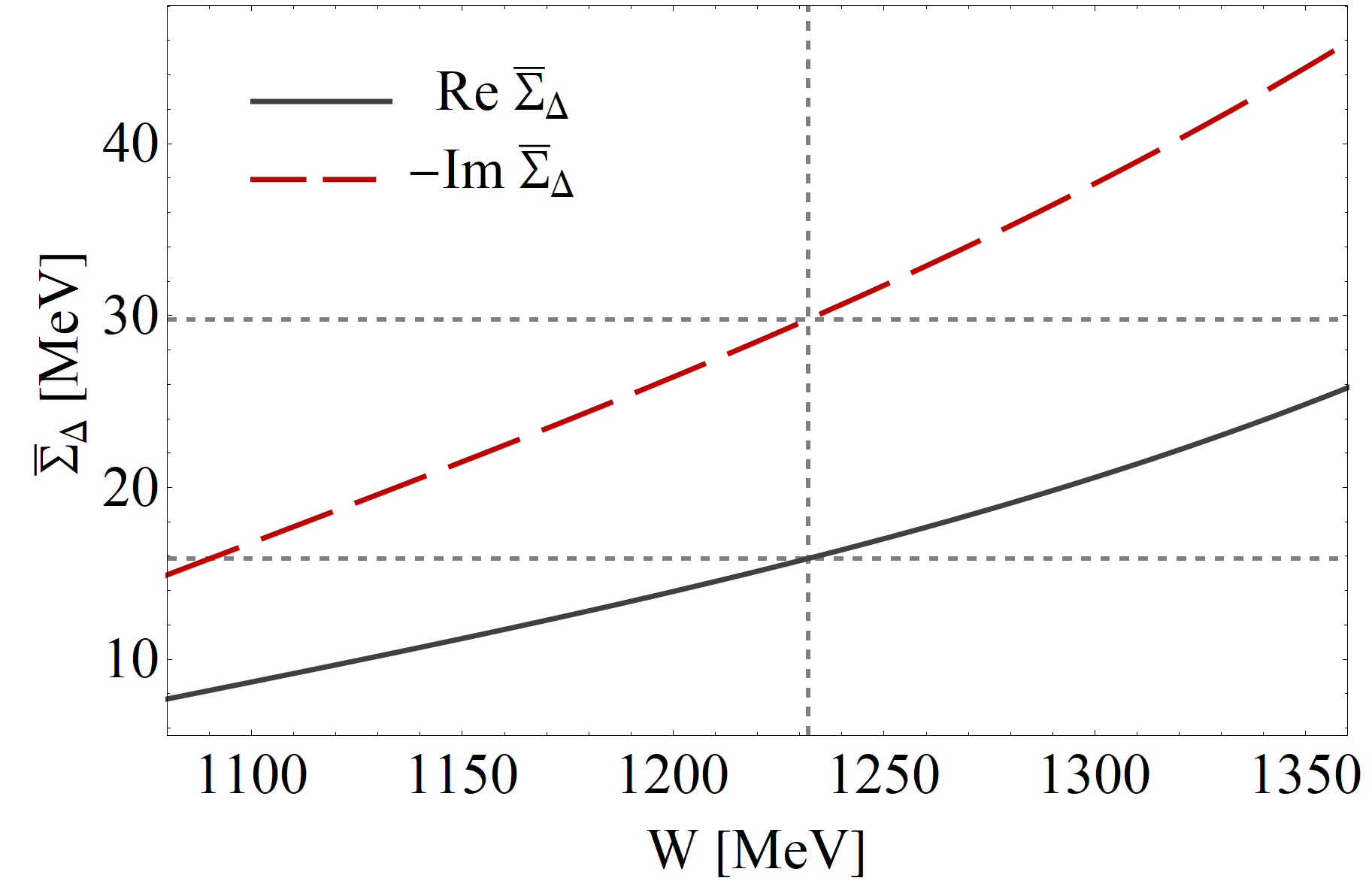}} 
\caption{
Upper panel:
the total and bare $\Delta$-resonance elastic decay widths, denoted as $\Gamma$ and $\Gamma_\Delta$ respectively, as functions of the pion-nucleon reaction energy $W$ obtained within R$\Delta$M.
Lower panel:
the effective total $\Delta$ self-energy $\overline \Sigma_\Delta$ as a function of $W$, Eq.~(\ref{barSigma(W)}),  with $\Sigma_\Delta = (12.9- i \, 33.2)\SI{ }{MeV}$.
The short dashed gray lines represent the resonance position and the magnitude of the corresponding quantities.
 }
\label{fig:bar-sigma}
\end{figure}

The $\Delta$ self-energy $\Sigma_\Delta$ in Eq.~(\ref{K-in-medium}) is not yet the same as $\overline \Sigma_\Delta$ in Eq.~(\ref{bar-G-modified}).
As was discussed above, the quantities entering Eq.~(\ref{bar-G-modified}) incorporate the interference of the bare resonant with the background corrections.
Within R$\Delta$M this interference explicitly arises in the scattering amplitude upon substituting the $K$-matrices, Eqs.~(\ref{K-in-medium}) and~(\ref{K-background}), into Eq.~(\ref{f(K)}).
The resulting $P_{33}$ channel pion-nucleon scattering amplitude can be expressed in the form
\be
f^1_{33} = \frac1{k_0} \frac{\Gamma(W)/2}{m_\Delta - W - i\, \Gamma(W)/2 + \overline \Sigma_\Delta(W)},
\label{f33(Sigma)}
\ee
where the total elastic width is
\be
\Gamma(W) = \Gamma_\Delta(W) + 2(m_\Delta - W) k_0 K^{1 (B)}_{33}
\ee
and we have introduced the effective total $\Delta$ self-energy 
\be
\overline \Sigma_\Delta(W) = \Sigma_\Delta \frac{\Gamma_\Delta(W)}{\Gamma(W) + 2 \Sigma_\Delta k_0 K^{1 (B)}_{33}}.
\label{barSigma(W)}
\ee
In this way, the total elastic width and total $\Delta$ self-energy, $\Gamma(W)$ and $\overline \Sigma_\Delta(W)$, differ from the corresponding bare-resonance quantities, $\Gamma_\Delta(W)$ and $\Sigma_\Delta$,  due to the inclusion of non-resonant contributions.
As follows from Eq.~(\ref{barSigma(W)}), the influence of the background brings energy dependence to the effective $\Delta$ self-energy in Eq.~(\ref{f33(Sigma)}).
The upper and lower panels of Fig.~\ref{fig:bar-sigma} illustrate the total and bare-resonance $\Delta$ decay widths, as well as the total $\Delta$ self-energy provided by the R$\Delta$M, respectively.

\begin{figure}[!thb]
\center{\includegraphics[width=0.99\linewidth]{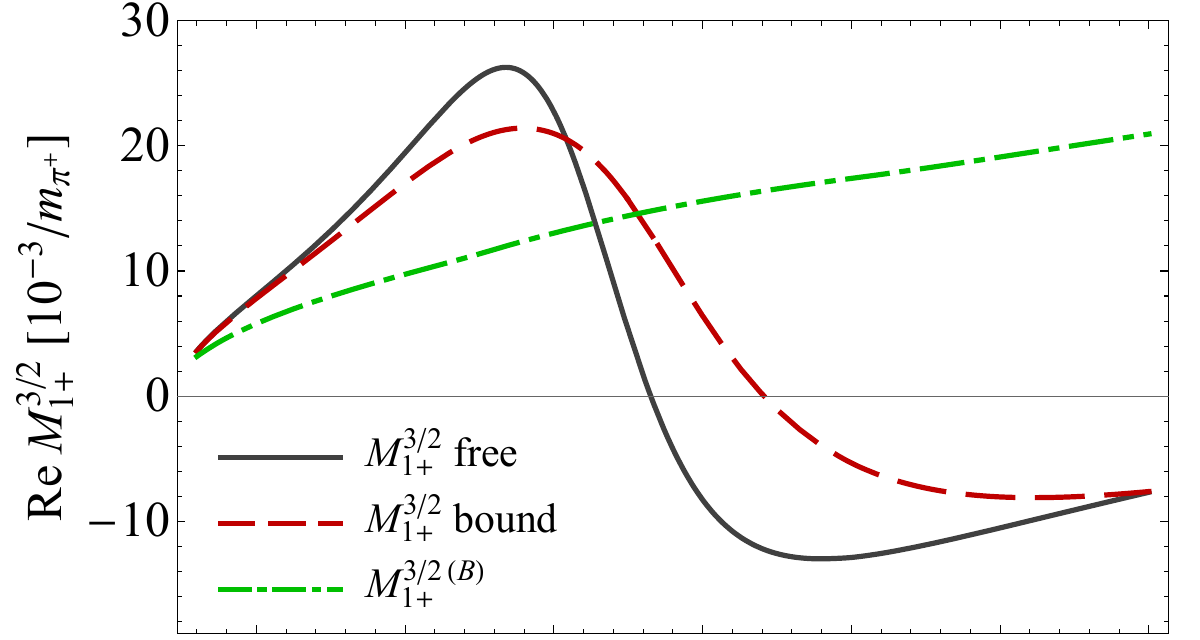}
\includegraphics[width=0.99\linewidth]{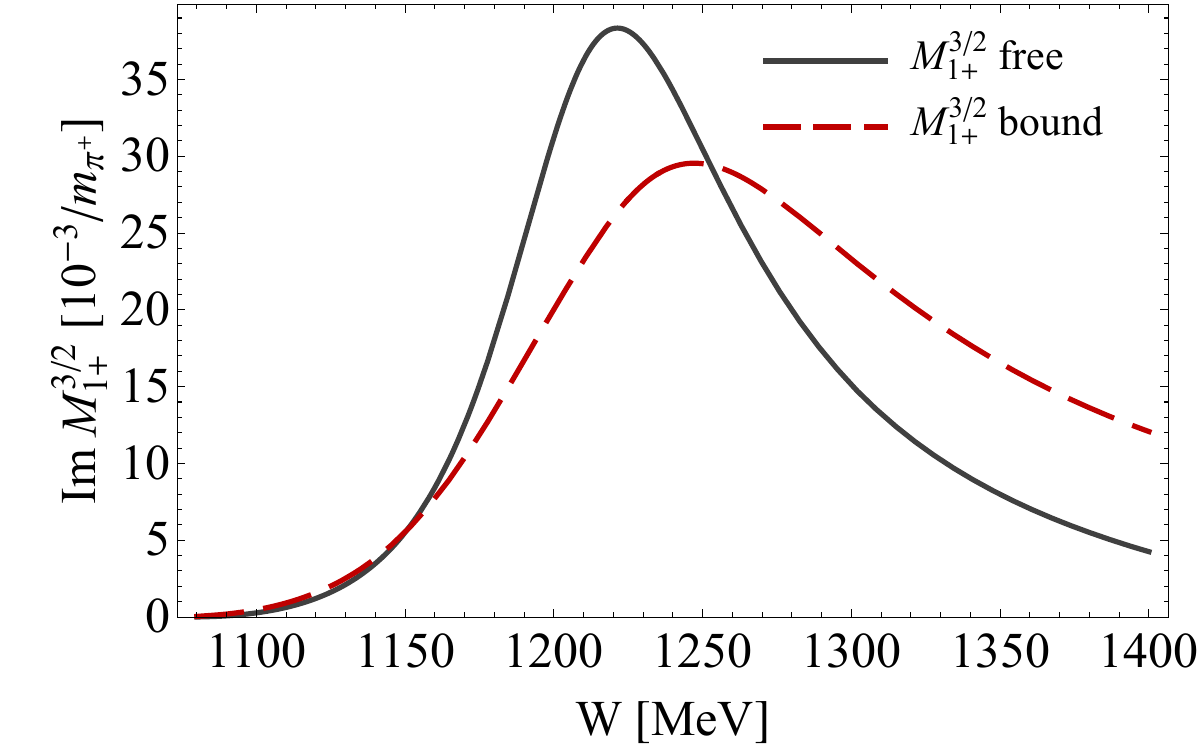}} 
\caption{
The real and imaginary parts of the $M_{1+}^{3/2}$
multipole. 
The solid black curves correspond to the amplitude in free space given by MAID98~\cite{drechsel2007unitary}. 
The dashed red curves represent the modified multipole in the nuclear medium as given by Eq.~(\ref{bar-G-modified}) with the effective $\Delta$ self-energy illustrated in Fig.~\ref{fig:bar-sigma}.
The dash-dotted green curve is the non-resonant background contribution in MAID98~\cite{drechsel1999unitary}.
}
\label{fig:M1p3-free-bound}
\end{figure}

Finally, we utilize $\overline \Sigma_\Delta$ given by Eq.~(\ref{barSigma(W)}) to calculate the pion photoproduction amplitude in the nuclear medium using Eq.~(\ref{bar-G-modified}).
The modified $M_{1+}^{3/2}$ multipole resulting from this process is compared with its free-space counterpart in Fig.~\ref{fig:M1p3-free-bound}.
The plot illustrates that the medium effects significantly modify the magnitude and resonance position of the amplitude.
The MAID98 background contribution, $M_{1+}^{3/2(B)}$, also shown in the figure, exhibits a monotonically increasing trend with energy.
This behavior mainly arises from vector mesons other than $\rho$ and $\omega$, which are not accounted for in the MAID model and affect the results at higher energies above the $\Delta(1232)$ resonance.

In contrast to our model, which assumes that $\overline \Sigma_\Delta$ is momentum-independent, Ref.~\cite{Drechsel:1999vh} introduces dependence on the momentum transfer $q$ into $\overline \Sigma_\Delta$ through the factor $F(q) = \exp(-\beta q^2)$. 
Here, $F(q)$ is the $s$-shell harmonic oscillator form factor of ${}^4$He with $\beta = \SI{0.54}{fm^2}$. This phenomenological modification is assumed to be universal and applicable to heavier nuclei with the same $\beta$.
We do not include this factor in $\overline \Sigma_\Delta$ to maintain consistency with our model for pion scattering. 
Incorporating this factor has been observed to spoil the fitting of the scattering data, particularly the differential elastic cross sections at large angles.

\section{The second-order potential for nuclear pion photoproduction }
\label{sec:Vgamma}

In this section, we derive the explicit form of the $\pi^0$ photoproduction potential in momentum space, as given by Eq.~(\ref{V_gamma-def}), for spin- and isospin-zero nuclei.
The derivation closely aligns with the development of the pion-nucleus scattering potential presented in Ref.~\cite{Tsaran:2023qjx}.

According to Eq.~(\ref{Ugamma-2term}), the potential is approximated as the sum of two terms:
\be
V_\gamma(\bm k', \bm k) \approx V_\gamma^{(1)}(\bm k', \bm k) + V_\gamma^{(2)}(\bm k', \bm k),
\ee
where, within the impulse approximation, the first-order part of the potential is
\be
V_\gamma^{(1)}(\bm k', \bm k) = - \frac{\sqrt{\mathscr{M}(k') \mathscr{M}_\gamma( k)}}{2\pi} A \langle \pi(\bm k'), \Psi_0 | \hat t^\gamma | \gamma(\bm k), \Psi_0 \rangle
\label{Ugamma1st-def}
\ee
and the second-order part is given by
\begin{multline}
V_\gamma^{(2)}(\bm k', \bm k) = - \frac{\sqrt{\mathscr{M}(k') \mathscr{M}_\gamma( k)}}{2\pi} \\
\times A(A - 1)  \langle \pi(\bm k'), \Psi_0 | \hat t_2 \hat G \hat P_\emptyset \hat t^\gamma_1 | \gamma(\bm k), \Psi_0 \rangle
\label{Ugamma2nd-def}.
\end{multline}

Using the matrix element given by Eq.~(\ref{<tgamma>-fgamma()}) and \textit{the optimal factorization approximation} arising from Eqs.~(\ref{p_eff-def}), we obtain the following first-order spin-isospin-averaged photoproduction potential in momentum space:
\begin{multline}
V_\gamma^{(1)}(\bm k', \bm k) = \mathscr{W}_\gamma(\bm k', \bm k) 
\\
\times
\frac{[\bm k'_\text{2cm} \times \bm k_\text{2cm}^{\vphantom{\prime}}] \cdot \bm \epsilon^\lambda}{k'_\text{2cm} k_\text{2cm}^{\vphantom{\prime}}} 
F_2^+(\bm k'_\text{2cm}, \bm k_\text{2cm}^{\vphantom{\prime}}) \rho(\bm k' - \bm k),
\label{Vgamma(1)}
\end{multline}
with the phase space factor
\be
\mathscr{W}_\gamma(\bm k', \bm k) = \sqrt{\frac{\mathscr{M}(k') \mathscr{M}_\gamma(k)}{\mu(\bm k', \bm p_N')\mu_\gamma(\bm k, \bm p_N)}},
\ee
and the nuclear form factor $\rho(q)$, normalized to $\rho(0) = A$.
In this work, the numerical values of $\rho(q)$ are derived from the nuclear charge form factors provided by the model-independent Fourier-Bessel analyses (Refs.~\cite{Cardman:1980dja} for ${}^{12}$C and~\cite{DeJager:1987qc} for ${}^{40}$Ca), corrected for the electromagnetic form factors of proton and neutron.
The nucleon form factors are taken from the global fits of electron scattering data of Ref.~\cite{Ye:2017gyb}.

The second-order component of the potential characterizes the subsequent rescattering of the generated pion on a second nucleon, occurring concurrently with the excitation and de-excitation processes within the nucleus.
To calculate the matrix element in Eq.~(\ref{Ugamma2nd-def}), we need to account for the spin-isospin structure of both $\hat t^\gamma$  and $\hat t$.
While the photoproduction amplitude is decomposed according to Eqs.~(\ref{CGLN-def}) and~(\ref{photo-isospin-struct}), the scattering amplitude is commonly presented as~\cite{Ericson:1988gk}
\be
\hat t = \hat t^{(0)} + \hat t^{(1)} \ \hat{\bm t} \cdot \hat{\pmb \tau} + 
\left(\hat t^{(2)} +
\hat t^{(3)} \ \hat{\bm t} \cdot \hat{\pmb \tau} \right) \hat{\pmb \sigma} \cdot \bm n,
\label{t-isospin-struct}
\ee
where $\hat{\bm t}$ is the pion isospin operator, and $\bm n$ is the normal to the pion-nucleon scattering plane.

The spin and isospin dependence in Eq.~(\ref{Ugamma2nd-def}) can be explicitly factored out.
For spin-zero nuclei, the direct summation over spin components yields:

\begin{widetext}
\begin{multline}
\sum\limits_{s,s^\prime = -1/2}^{1/2} \chi_1^\dag(s) \chi_2^\dag(s^\prime)  \left(\hat t^{(0,1)} + \hat t^{(2,3)} \hat{\pmb \sigma}_2 \cdot \bm n_2 \right) \hat f^\gamma_1(\bm k''_\text{2cm}, \bm k_\text{2cm}^{\vphantom{\prime}}) \chi_1(s^\prime) \chi_2(s) =  \\
2 \left( \boldsymbol \epsilon^\lambda \cdot \bm n_1 \hat F_2(\bm k''_\text{2cm}, \bm k_\text{2cm}^{\vphantom{\prime}}) \hat t^{(0,1)}
+ i \boldsymbol \epsilon^\lambda \cdot \bm n_2 \hat{\mathscr{F}}_1(\bm k''_\text{2cm}, \bm k_\text{2cm}^{\vphantom{\prime}}) \hat t^{(2,3)}
+ i \frac{(\bm k_\text{2cm}^{\vphantom{\prime}} \cdot \bm n_2) (\bm k_\text{2cm}'' \cdot \boldsymbol \epsilon^\lambda)}{k_\text{2cm}^{\vphantom{\prime}} k_\text{2cm}''}
\hat{\mathscr{F}}_3(\bm k''_\text{2cm}, \bm k_\text{2cm}^{\vphantom{\prime}}) \hat t^{(2,3)}
\right),
\label{Ugamma2-spin-struct}
\end{multline}
where $\chi_i(s)$ is the nucleon spinor, $\bm n_1 = [\bm k_\text{2cm}'' \times \bm k_\text{2cm}^{\vphantom{\prime}}] / (k_\text{2cm}'' k_\text{2cm}^{\vphantom{\prime}})$ and $\bm n_2 = [\bm k_{\text{2cm}'}' \times \bm k_{\text{2cm}'}''] / ( k_{\text{2cm}'}' k_{\text{2cm}'}'')$.
The subscript "$\text{2cm}'$" corresponds to the c.m. frame of the pion and the second nucleon.  
The spin-flip pion photoproduction amplitudes $\hat{\mathscr{F}}_{1,3}$ amplitudes are related to the CGLN amplitudes as
\begin{subequations}
\begin{align}
&\hat{\mathscr{F}}_1(\bm k', \bm k) = \hat F_1(\bm k', \bm k) - \frac{\bm k' \cdot \bm k}{k' k} \hat F_2(\bm k', \bm k), \\
&\hat{\mathscr{F}}_3(\bm k', \bm k) = \hat F_2(\bm k', \bm k) + \hat F_3(\bm k', \bm k) .
\end{align}
\end{subequations}
Note that contributions proportional to $\hat F_4 \hat t^{(2)}$ exactly cancel.
This cancellation arises from the fact that the operator $(\bm n \cdot \hat{\boldsymbol \sigma})$  serves as a spin projector onto the $\bm n$ axis.
Consequently, operators $(\bm n_2 \cdot \hat{\boldsymbol \sigma})$ and $(\bm k' \cdot \hat{\boldsymbol \sigma})$ project onto perpendicular directions. 
Although these operators correspond to different nucleons, the summation over spin makes the final result proportional to $\bm n_2 \cdot \bm k' = 0$.

It is convenient to use the relation
\be
\langle \pi_b | \hat{\bm t} \cdot \hat{\pmb \tau} | \pi_a \rangle = i \varepsilon_{abc} \hat{\pmb \tau}_c = - \frac12 [\hat{\pmb \tau}_b, \hat{\pmb \tau}_a]
\ee
to write the isospin decomposition of Eq.~(\ref{photo-isospin-struct}) in a form similar to Eq.~(\ref{t-isospin-struct}).
Taking the sum over the isospin in Eq.~(\ref{Ugamma2nd-def}) for nuclei with zero isospin, one obtains
\begin{multline}
\sum\limits_{\tau,\tau^\prime = -1/2}^{1/2} \eta_1^\dag(\tau) \eta_2^\dag(\tau^\prime)  \left[
\delta_{b a} \hat t^{(0,2)} - \frac12 [\hat{\pmb \tau}_a, \hat{\pmb \tau}_b] \hat t^{(1,3)}
\right]_2
\left[ 
\delta_{b 3} A_{l\pm}^+ + \frac12 [\hat{\pmb \tau}_b, \hat{\pmb \tau}_3] A_{l\pm}^- + \hat{\pmb \tau}_b A_{l\pm}^0
\right]_1
\eta_1(\tau^\prime) \eta_2(\tau) \\
=  
2 \left( 
\hat t^{(0,2)} \hat A^+ - 2 \, \hat t^{(1,3)} \hat A^-
\right),
\label{Ugamma2-isospin-struct}
\end{multline}
where $\eta_i(\tau)$ is the nucleon isospinor and the subscript $i=1,2$ in $[\cdots]_i$ refers to the nucleon index as in Eq.~(\ref{Ugamma2nd-def}).

Finally, neglecting the nuclear excitation energies in $\hat G$ and combining Eqs.~(\ref{Ugamma2-spin-struct}) and~(\ref{Ugamma2-isospin-struct}), the second-order part of the potential in momentum space for spin- and isospin-zero nuclei can be written as
\begin{multline}
V_\gamma^{(2)}(\bm k' ,\bm k) 
= \int \frac{\diff \bm k''}{2\pi^2}   \frac{\mathscr{W}(\bm k', \bm k'') \mathscr{W}_\gamma(\bm k'', \bm k)  }{k_0^2 - {k''}^2+ i\varepsilon} \left[ 
f^{(0)}(\bm k^{\prime} ,\bm k^{\prime\prime}) 
f^{(0)}_\gamma(\bm k^{\prime\prime} ,\bm k) 
C_0(\bm k^\prime - \bm k^{\prime\prime},  \bm k^{\prime\prime} - \bm k)
\right.\\
\left.
+
\left(
2 f^{(1)}(\bm k^{\prime} ,\bm k^{\prime\prime}) 
f^{(1)}_\gamma(\bm k^{\prime\prime} ,\bm k) 
+
f^{(2)}(\bm k^{\prime} ,\bm k^{\prime\prime}) 
f^{(2)}_\gamma(\bm k^{\prime\prime} ,\bm k) 
+
2 f^{(3)}(\bm k^{\prime} ,\bm k^{\prime\prime}) 
f^{(3)}_\gamma(\bm k^{\prime\prime} ,\bm k) 
\right)
C_\text{ex}(\bm k^\prime - \bm k^{\prime\prime},  \bm k^{\prime\prime} - \bm k)
\right],
\label{V2_gamma-final}
\end{multline}
where 
\begin{subequations}
\begin{align}
&f^{(0,1)}_\gamma(\bm k^{\prime\prime} ,\bm k) = \pm \boldsymbol \epsilon^\lambda \cdot 
\bm n_1 F_2^\pm(\bm k''_\text{2cm}, \bm k_\text{2cm}^{\vphantom{\prime}}),
\\
&f^{(2,3)}_\gamma(\bm k^{\prime\prime} ,\bm k) = \pm i \boldsymbol \epsilon^\lambda \cdot \bm n_2 \mathscr{F}_1^\pm(\bm k''_\text{2cm}, \bm k_\text{2cm}^{\vphantom{\prime}})
\pm 
i \frac{(\bm k_\text{2cm} \cdot \bm n_2) (\bm k_\text{2cm}'' \cdot \boldsymbol \epsilon^\lambda)}{k_\text{2cm}^{\vphantom{\prime}} k_\text{2cm}''} 
\mathscr{F}_3^\pm(\bm k''_\text{2cm}, \bm k_\text{2cm}^{\vphantom{\prime}}).
\end{align}
\end{subequations}
with $\mathscr{F}_{1,3}^\pm$ holding the same meaning as defined in Eq.~(\ref{photo-isospin-struct}). 
The scattering amplitudes $f^{(i)}$ are the same as for the scattering potential: 
\begin{subequations}
\begin{align}
&f^{(0,1)}(\bm k^{\prime} ,\bm k^{\prime\prime}) = 
\left(b_{0,1} + c_{0,1}  \, \bm k'_{\text{2cm}'} \cdot \bm k''_{\text{2cm}'} \right)  {v}(k') {v}(k), \\
&f^{(2,3)}(\bm k^{\prime} ,\bm k^{\prime\prime}) = i \, s_{0,1} \, k'_{\text{2cm}'}   k''_{\text{2cm}'}  {v}(k') {v}(k),
\end{align}
\end{subequations}
where $b_{0,1}$, $c_{0,1}$, and $s_{0,1}$ are the standard scattering parameters~\cite{Ericson:1988gk}. 
The $P_{33}$ channel contribution to $p$-wave parameters $c_{0,1}$ and $s_{0,1}$ is modified in nuclear medium as described in Sec.~\ref{sec:M1p3-in-medium}.

The properties of the nucleon distribution contribute to the second-order potential via the correlation functions $C_0$ and $C_\text{ex}$, which are defined in coordinate space as
\begin{subequations}
\begin{align}
&C_\text{ex}(\bm r_1, \bm r_2) = \rho(\bm r_1) \rho(\bm r_2) - \rho_2(\bm r_1, \bm r_2), 
\label{rho_ex(r)-def}\\
&C_0(\bm r_1, \bm r_2) = 
C_\text{ex}(\bm r_1, \bm r_2) - \frac1A \rho(\bm r_1) \rho(\bm r_2),
\label{C(r)-def}
\end{align}
\label{rho_ex-and-C(r)}%
\end{subequations}
where $\rho(\bm r)$ and $\rho_2(\bm r_1, \bm r_2)$ are the one- and two-body nuclear density distributions, respectively.
The corresponding quantities in momentum space are obtained by the Fourier transform (see Ref.~\cite{Tsaran:2023qjx} for details).
In our calculations, we employ the Slater determinant form of the total nuclear wave function, with single-particle wave functions of nucleons derived from the harmonic oscillator shell model. 
The resulting correlation functions take the following form for ${}^{12}$C: 
\begin{subequations}
\begin{align}
&C_\text{ex}(\bm q_1, \bm q_2) = \left( 12 - \frac43 a^2 (q_1^2 + q_2^2) -
  4 \sqrt{\frac23} a^2 \bm q_1 \cdot \bm q_2 + \frac23 a^4 (\bm q_1 \cdot \bm q_2)^2 \right)
  \exp\left[ - \frac14 \frac{A-1}A a^2 \left(q_1^2 + q_2^2 \right) \right],\\
&C_0(\bm q_1, \bm q_2) = \left(
  -  4 \sqrt{\frac23} a^2 \bm q_1 \cdot \bm q_2 + \frac23 a^4 (\bm q_1 \cdot \bm q_2)^2  - \frac4{27} a^4 q_1^2 q_2^2 \right)
  \exp\left[ - \frac14 \frac{A-1}A  a^2 \left(q_1^2 + q_2^2 \right) \right],
\end{align}
\label{C-D-12C}%
\end{subequations}
and for ${}^{40}\text{Ca}$:
\begin{subequations}
\begin{align}
&C_\text{ex}(\bm q_1, \bm q_2) = \left( 40 - 10 a^2 (\bm q_1 + \bm q_2)^2 + \frac12 a^4 \left((\bm q_1 + \bm q_2)^4 + 10 (\bm q_1 \cdot \bm q_2)^2 \right) 
- \frac12 a^6 (\bm q_1 \cdot \bm q_2)^2 \left(q_1^2 + q_2^2 + \bm q_1 \cdot \bm q_2 \right) 
\right.  \notag\\  
&\qquad\qquad\qquad\left.
+ \frac1{16} a^8 (\bm q_1 \cdot \bm q_2)^4
\right)
  \exp\left[ - \frac14 \frac{A-1}A a^2 \left(q_1^2 + q_2^2 \right) \right],
  \\
&C_0(\bm q_1, \bm q_2) = \left(
    -20 a^2 \bm q_1 \cdot \bm q_2 + \frac12 a^4 
    \left( 4 (\bm q_1 + \bm q_2)^2 \bm q_1 \cdot \bm q_2 + 6 (\bm q_1 \cdot \bm q_2)^2 - 3 q_1^2 q_2^2  \right)
+ \frac1{160} a^8 \left( 10 (\bm q_1 \cdot \bm q_2)^4 - q_1^4 q_2^4
\right)
\right.  \notag\\  
&\qquad\qquad\qquad\left.
- \frac1{8} a^6 \left( 4 ( q_1^2 + q_2^2 + \bm q_1 \cdot \bm q_2) (\bm q_1 \cdot \bm q_2)^2 - (q_1^2 + q_2^2) q_1^2 q_2^2
\right)
\right)
\exp\left[ - \frac14 \frac{A-1}A a^2 \left(q_1^2 + q_2^2 \right) \right].
\end{align}
\label{C-D-40Ca}%
\end{subequations}
The harmonic oscillator parameter $a$ is $\SI{1.63}{fm}$ for ${}^{12}$C and $\SI{1.98}{fm}$ for ${}^{40}\text{Ca}$.

\end{widetext}

The second-order part of the photoproduction potential, as expressed in Eq.~(\ref{Ugamma2nd-def}), has a form similar to that of the scattering potential, and the four terms on the right-hand side carry the same physical meaning.
The first term describes spin-isospin averaged individual nucleon scattering on two nucleons.
The term proportional to $f^{(1)} f^{(1)}_\gamma$ ($\hat f^{(2)} \hat f^{(2)}_\gamma$) describes the contribution of the intermediate charge exchange (nucleon spin-flip) on two nucleons, keeping the scattered nucleus in the ground state.
Similarly, the term $\hat f^{(3)} \hat f^{(3)}_\gamma$ represents the simultaneous exchange of both spin and isospin.

\section{Results and discussion}
\label{sec:results-photo}

In this section, we compare the predictions of the described model for the coherent nuclear pion photoproduction on ${}^{12}$C and ${}^{40}$Ca with experimental data and theoretical model calculations derived from Ref.~\cite{Drechsel:1999vh}. 
Note that no fitting of the $\Delta$ self-energy was done, and the obtained results are based on the parameters from the multi-energy fit to $\pi^\pm$-${}^{12}$C scattering data in the energy range of 80–\SI{180}{MeV} pion laboratory kinetic energy.

The comparison data are sourced from Refs.~\cite{Krusche:2002iq} and~\cite{Tarbert:2007whk}.
These experiments were performed with the TAPS~\cite{Novotny:1992bw, Gabler:1994ay} and Crystal Ball~\cite{CrystalBall:2001uhc} detectors, respectively, at the electron microtron in Mainz (MAMI)~\cite{Kaiser:2008zza}, combining it with the Glasgow tagged photon beam~\cite{Anthony:1991eq, Hall:1996gh}.

\begin{figure}[!t]
\includegraphics[width=0.99\linewidth]{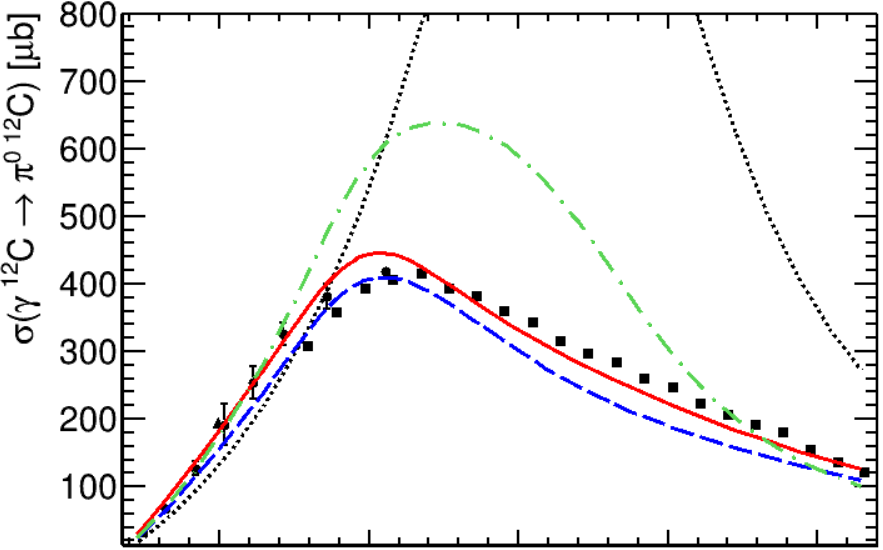}
\includegraphics[width=0.99\linewidth]{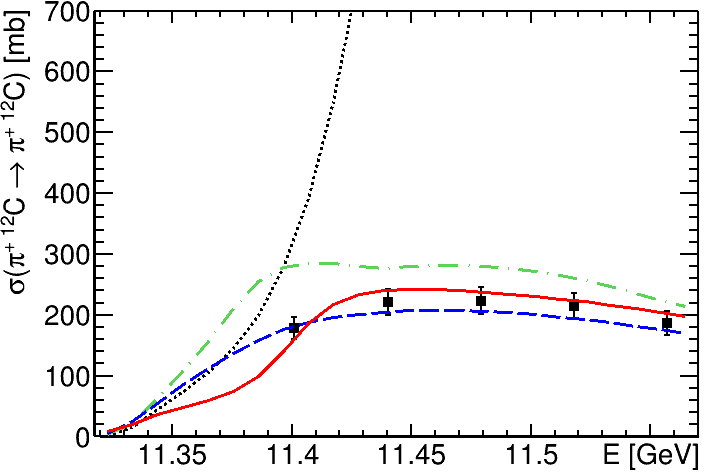}
\caption{
The integrated cross sections for coherent $\pi^0$ photoproduction (upper panel) and $\pi^+$ elastic scattering (lower panel) on ${}^{12}$C as functions of the total c.m. energy $E$ of the system.
The black dotted curves depict results under the PWIA. The green dot-dashed curves incorporate pion rescattering on the nucleus in its ground state (DWIA). 
The blue dashed curves are obtained when the in-medium modification of the elementary amplitudes is introduced to the DWIA. 
The red solid curves represent the complete calculation in which the second-order rescattering on excited nuclear states (middle diagram in Fig.~\ref{fig-Tgamma-diagrams}) is included.
The data on the upper panel are taken from Refs.~\cite{Tarbert:2007whk} ({\large $\bullet$}) \cite{Krusche:2002iq} ({\footnotesize $\blacksquare$}), and \cite{Gothe:1995zb} ($\blacktriangle$);
while data on the lower panel are from Ref.~\cite{Ashery:1981tq}.
}
\label{fig:scat-photo-corrections-effect}
\end{figure}

The unpolarized differential cross section for the nuclear pion photoproduction on a spin-zero target is expressed in terms of the nuclear photoproduction amplitude, Eq.~(\ref{Fgamma-final}), as
\be
\frac{d\sigma}{d \Omega}(\theta) = \frac{k_0}{k_0^\gamma} \frac12 \sum_{\lambda=\pm} \left| F_\gamma \right|^2.
\ee
The computation of the amplitude $F_\gamma$ involves several sequential steps, each depicted in the upper panel of Fig.~\ref{fig:scat-photo-corrections-effect}, showing the integrated cross section for the $\pi^0$ photoproduction on ${}^{12}$C in the energy range from threshold to photon laboratory kinetic energy of $\omega^\gamma_\text{lab} \approx \SI{400}{MeV}$.
Initially, we calculate the first-order momentum space photoproduction potential, Eq.~(\ref{Vgamma(1)}), with the free-space CGLN amplitude $F_2$.
This initial step is depicted by the dotted black curve in the plot assuming $F_\gamma \approx V_\gamma^{(1)}(\bm k', \bm k)$, which is known as the plane-wave impulse approximation (PWIA).
Subsequently, we incorporate the pion-nucleus rescattering as given by Eq.~(\ref{Fgamma-final}) with $V_\gamma(\bm k', \bm k) \approx V_\gamma^{(1)}(\bm k', \bm k)$ and utilize the pion-${}^{12}$C scattering amplitude $F(\bm k', \bm k)$ from Ref.~\cite{Tsaran:2023qjx}. 
This corresponds to the distorted wave impulse approximation (DWIA), represented by the dot-dashed green curve in Fig.~\ref{fig:scat-photo-corrections-effect}.
The next step, illustrated by the blue dashed curve in the plot, involves modification of the resonant $M_{1+}^{3/2}$ multipole in the nuclear medium, as given by Eq.~(\ref{bar-G-modified}), using the effective $\Delta$ self-energy $\overline \Sigma_\Delta$, Eq.~(\ref{barSigma(W)}).
Finally, the solid red curve represents the complete model incorporating the second-order photoproduction potential, $V_\gamma(\bm k', \bm k) \approx V_\gamma^{(1)}(\bm k', \bm k) + V_\gamma^{(2)}(\bm k', \bm k)$, utilizing the second-order part described by Eq.~(\ref{V2_gamma-final}). 
This last component accounts for the second-order rescattering on excited nuclear continuum states, involving intermediate nucleon spin-flip and charge exchange, corresponding to the middle diagram in Fig.~\ref{fig-Tgamma-diagrams}.

As seen from the upper panel of Fig.~\ref{fig:scat-photo-corrections-effect}, the DWIA prediction incorporating the effective $\Delta$ self-energy correction (the blue dashed curve) fails to match the magnitude of the data, although it captures the overall shape adequately.
Discrepancies between PWIA, DWIA, and the modification of the elementary photoproduction amplitude are minor at low incident photon energies but become significant in the $\Delta$-resonance region.
Incorporating the second-order component of the photoproduction potential, $V^{(2)}_\gamma$,  results in an upward shift of the predicted integrated cross section, increasing it by approximately 7\%  at $\omega^\gamma_\text{lab} \approx \SI{250}{MeV}$ to 17\% at $\omega^\gamma_\text{lab} \approx \SI{330}{MeV}$.
In the $\Delta$-resonance energy region, the inclusion of $V_\gamma^{(2)}$ into the calculation yields a relatively smaller effect compared to the influence of the final state interaction and the modification of the $\Delta$-resonance characteristics inside the nucleus.
However, the impact of $V^{(2)}_\gamma$ remains sizable both at low energies and in the $\Delta$-resonance region, providing the necessary correction for a satisfactory description of the data.

\begin{figure*}[!th]
\includegraphics[width=0.44\textwidth]{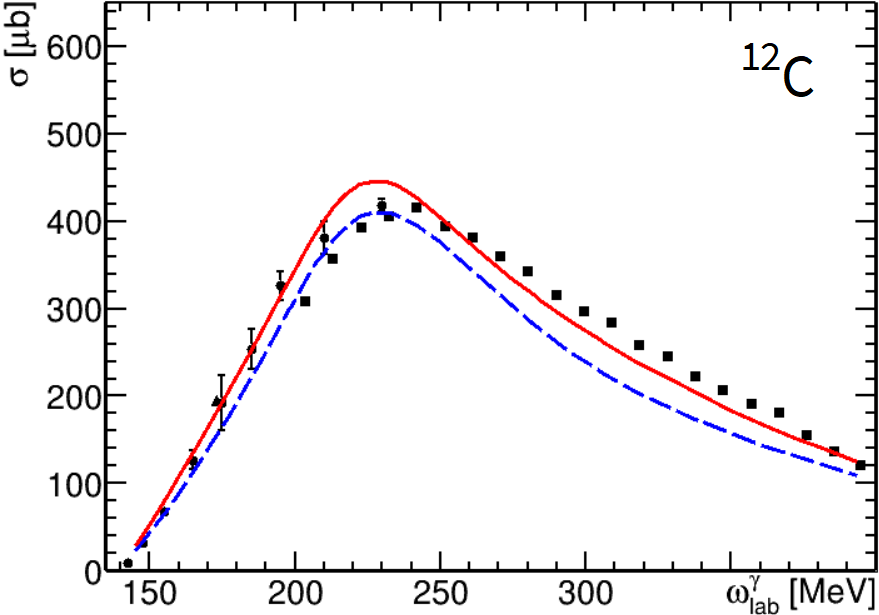}
\includegraphics[width=0.44\textwidth]{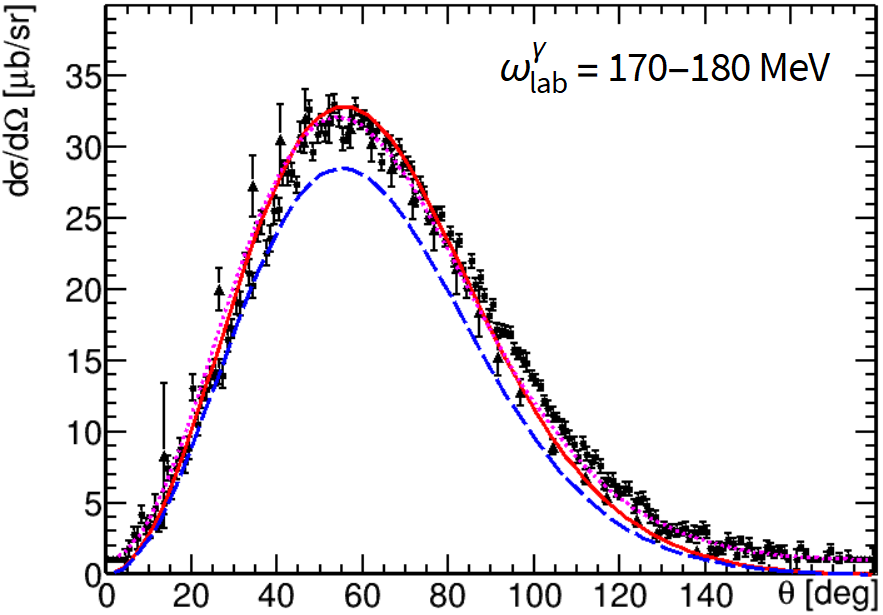}
\includegraphics[width=0.44\textwidth]{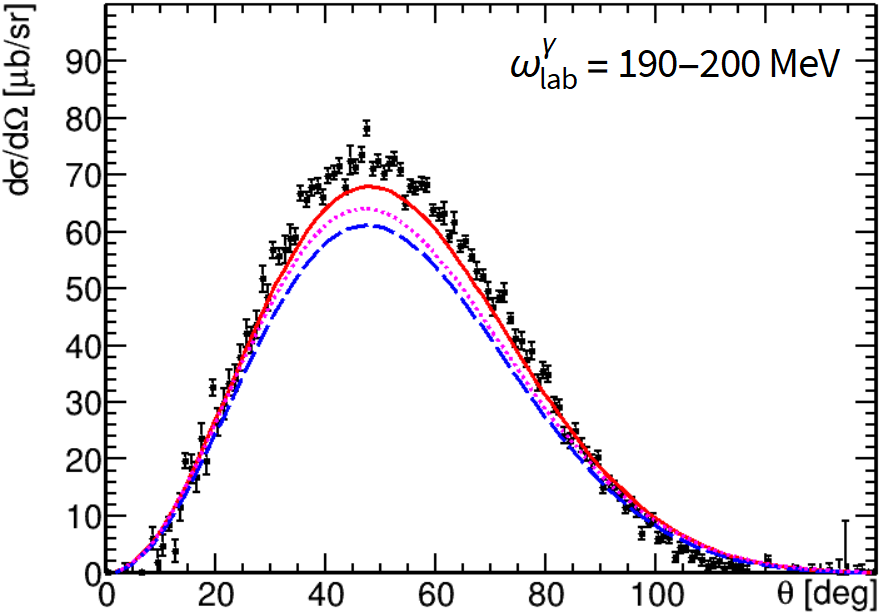}
\includegraphics[width=0.44\textwidth]{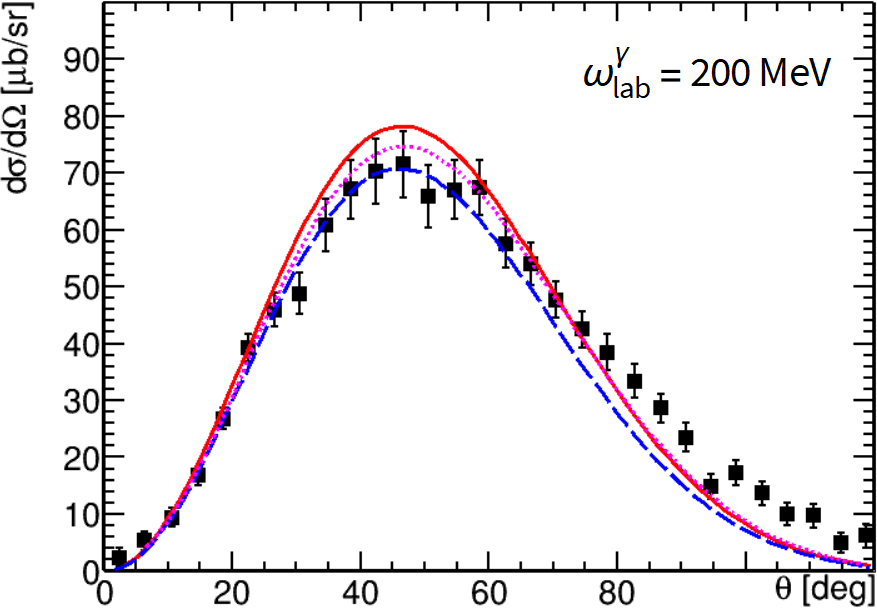}
\includegraphics[width=0.44\textwidth]{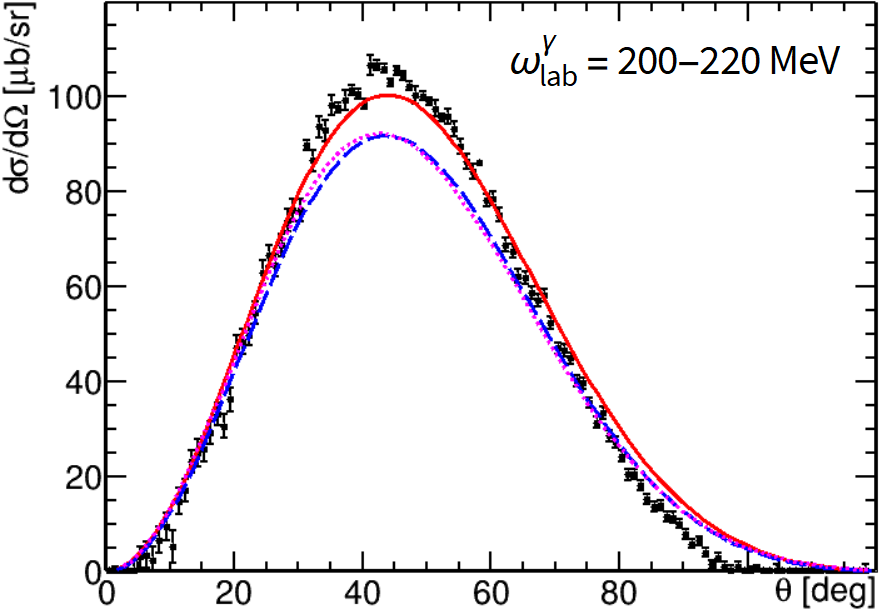}
\includegraphics[width=0.44\textwidth]{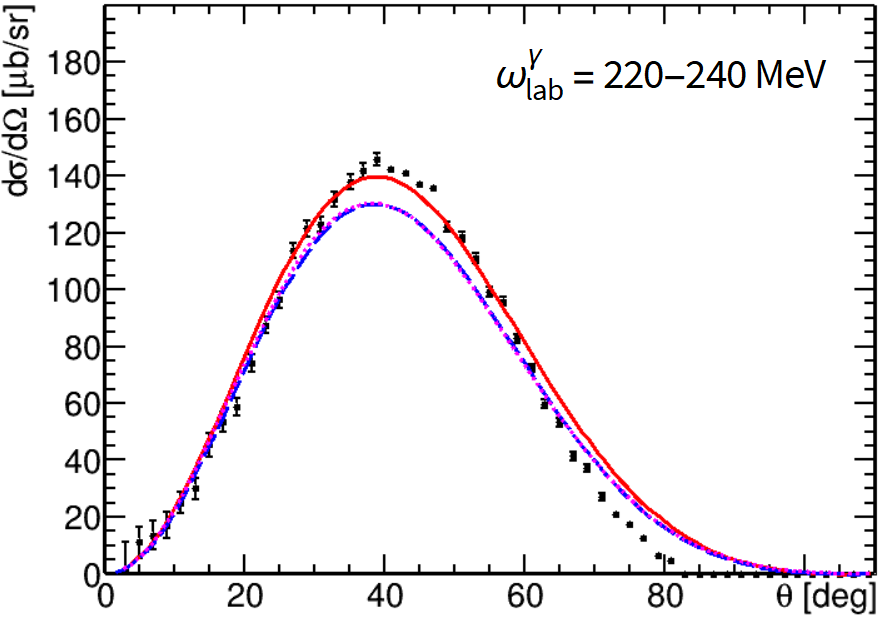}
\includegraphics[width=0.44\textwidth]{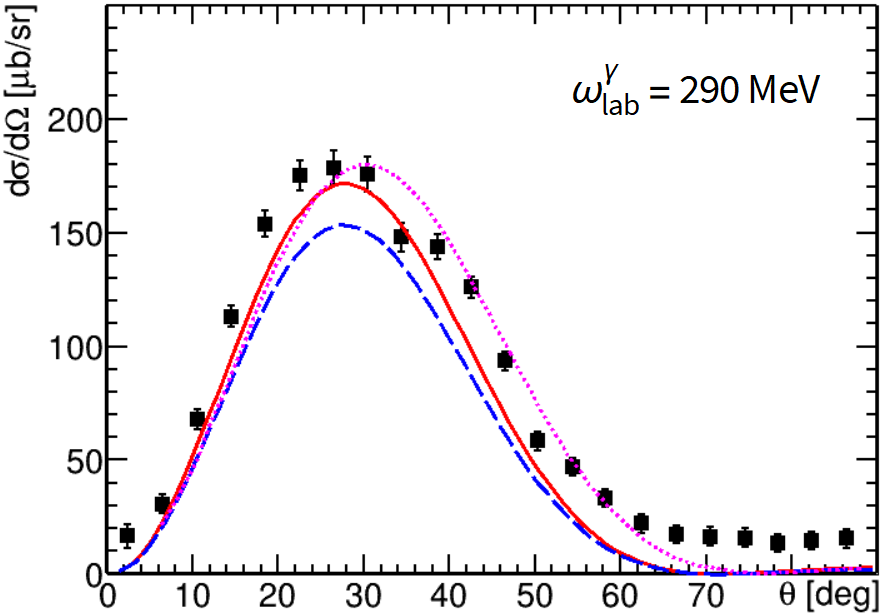}
\includegraphics[width=0.44\textwidth]{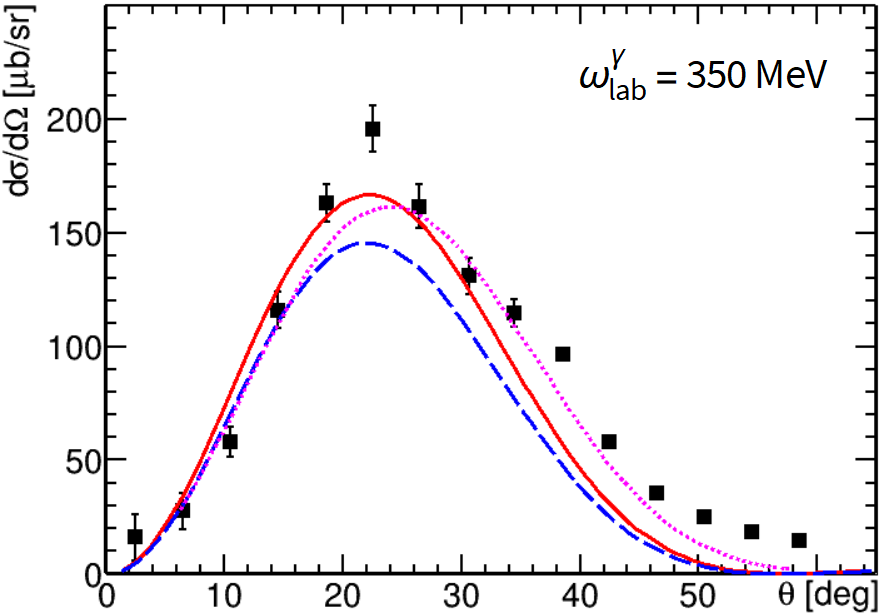}
\caption{
Comparison of theoretical predictions for the coherent $\pi^0$ photoproduction with the data for ${}^{12}$C.
The top left panel demonstrates the integrated cross section as a function of the photon laboratory energy $\omega_\text{lab}^\gamma$; the differential cross sections as functions of the scattering angle in the c.m. frame are shown in the other panels.
The solid red curves are the predictions of our model; the blue dashed curves are obtained by switching off the second-order part of the photoproduction potential, $V_\gamma^{(2)}$; the dotted magenta curves are results obtained with the theoretical model of Ref.~\cite{Drechsel:1999vh} according to Refs.~\cite{Tarbert:2007whk} and~\cite{Krusche:2002iq}. 
The experimental data for photon laboratory energies $\omega_\text{lab}^\gamma = 200, 290, \SI{350}{MeV}$ denoted by square markers are from Ref.~\cite{Krusche:2002iq}; the circles in the other panels represent data from Ref.~\cite{Tarbert:2007whk} measured for energy bins of 10 and $\SI{20}{MeV}$ width; the triangles on the upper right panel are from Ref.~\cite{Gothe:1995zb}, measured at $\omega_\text{lab}^\gamma = 170-\SI{177}{MeV}$.
}
\label{fig:12C-photo-res}
\end{figure*}

\begin{figure*}[!th]
\includegraphics[width=0.45\textwidth]{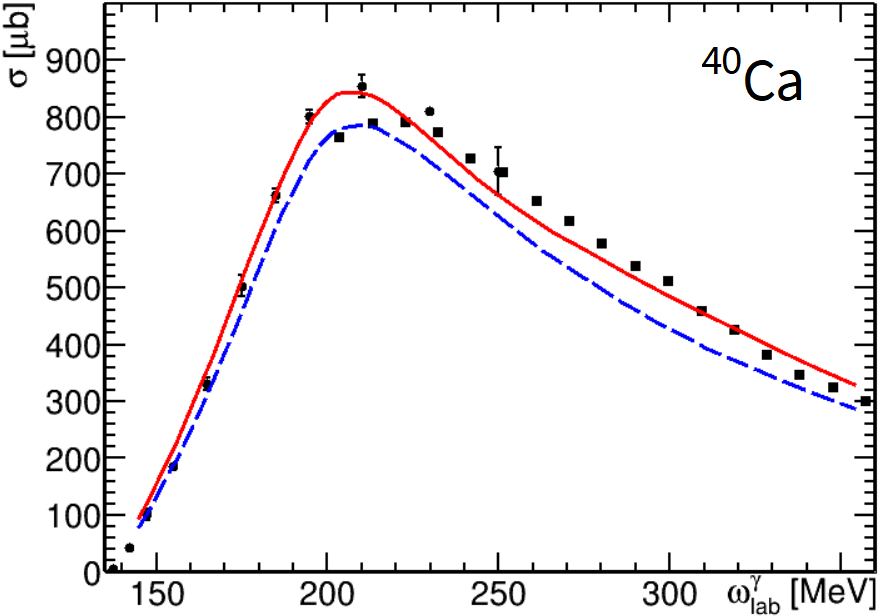}
\includegraphics[width=0.45\textwidth]{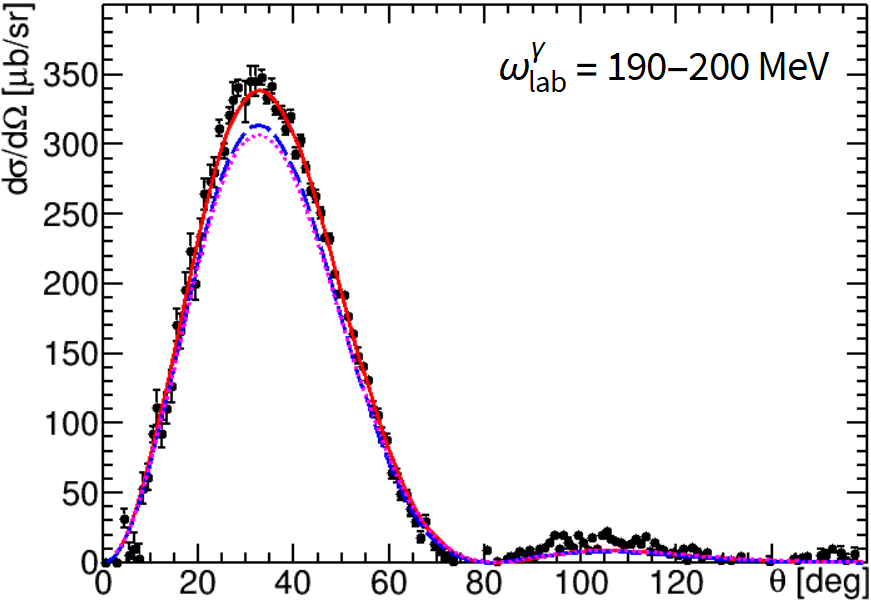}
\includegraphics[width=0.45\textwidth]{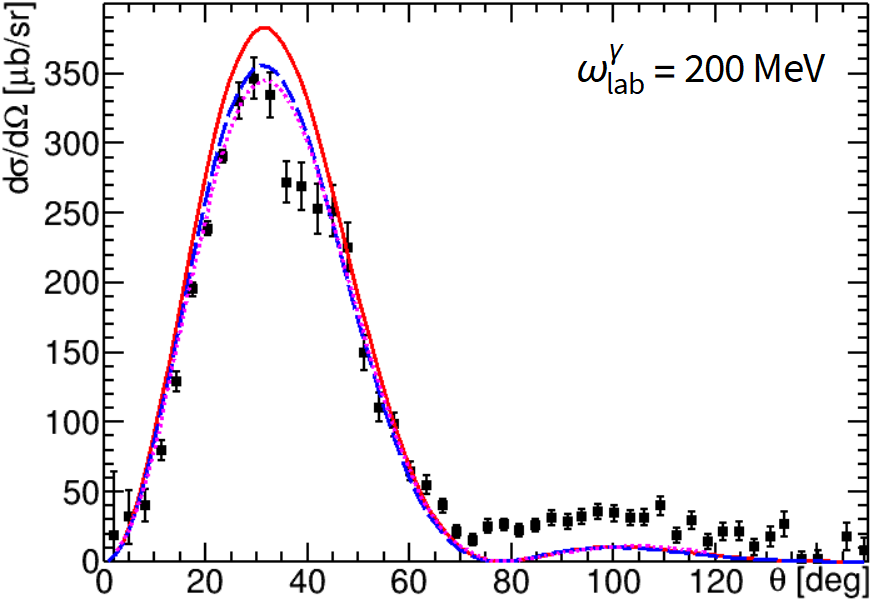}
\includegraphics[width=0.45\textwidth]{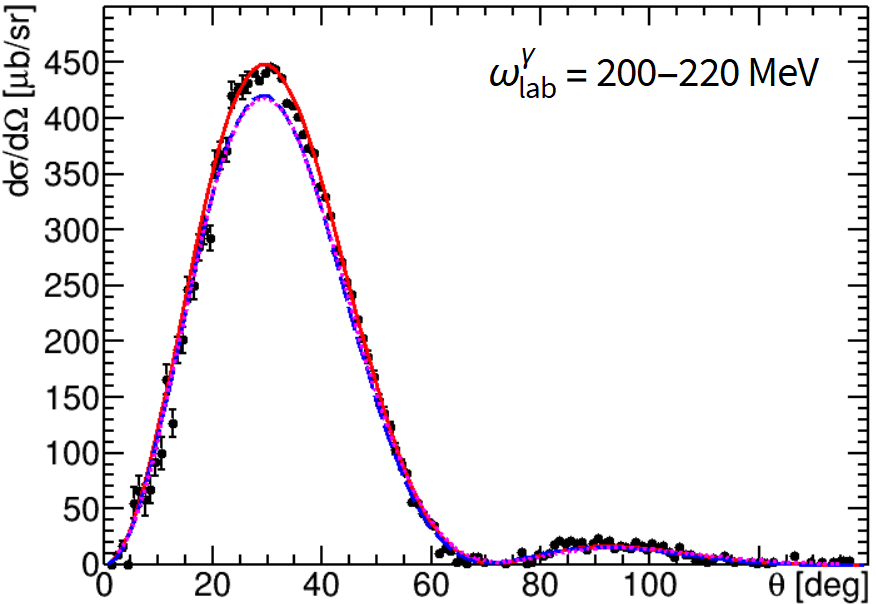}
\includegraphics[width=0.45\textwidth]{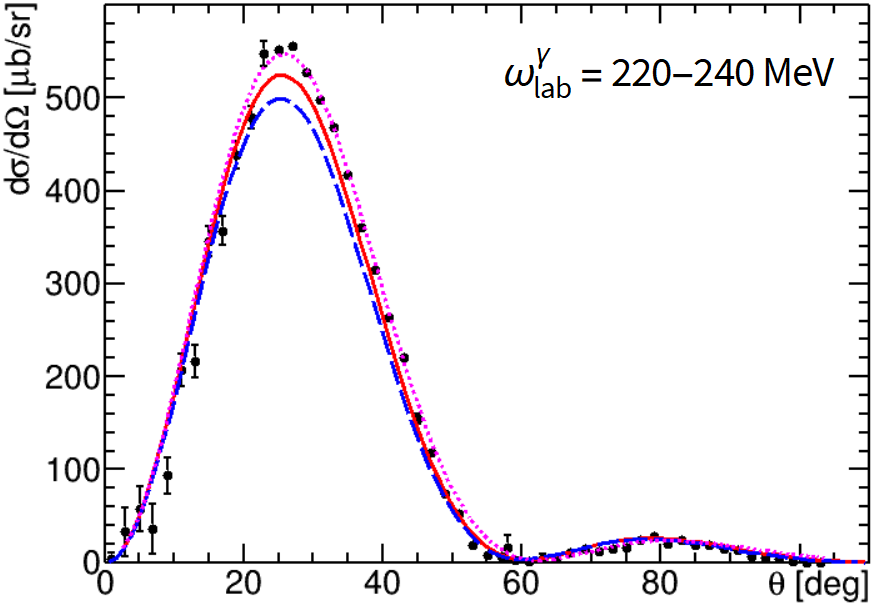}
\includegraphics[width=0.45\textwidth]{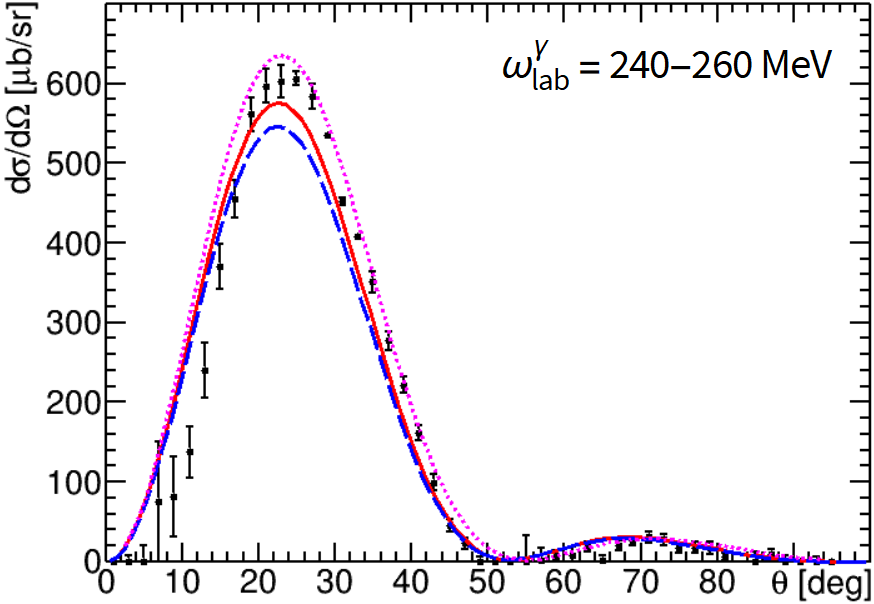}
\includegraphics[width=0.45\textwidth]{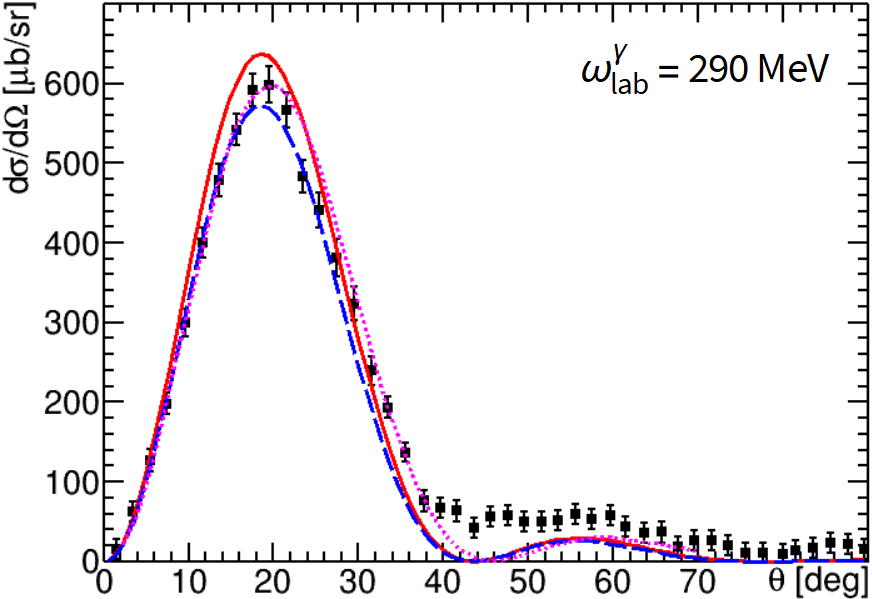}
\includegraphics[width=0.45\textwidth]{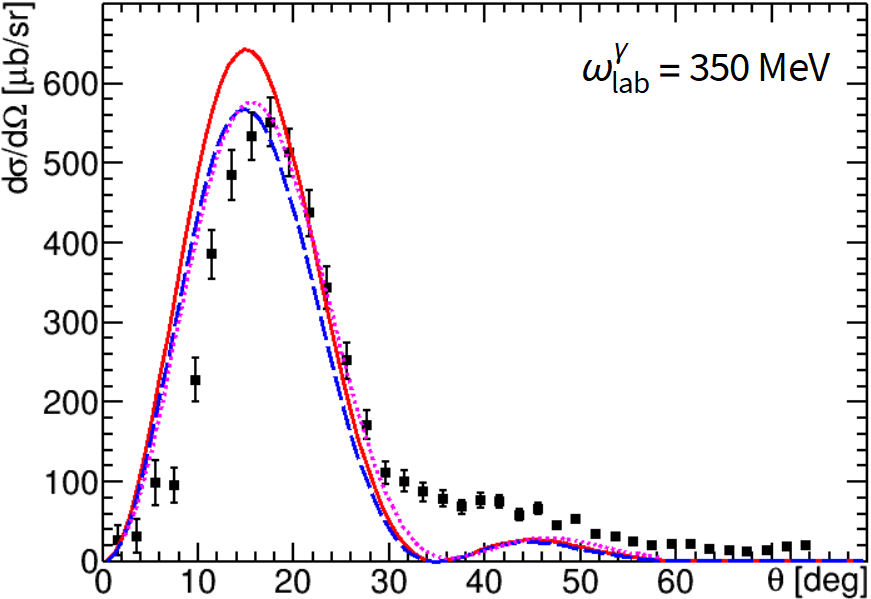}
\caption{
Comparison of theoretical predictions for the coherent $\pi^0$ photoproduction with the data for ${}^{40}$Ca.
The top left panel demonstrates the integrated cross section as a function of the photon laboratory energy $\omega_\text{lab}^\gamma$; the differential cross sections as functions of the scattering angle in the c.m. frame are shown in the other panels.
The data are from Refs.~\cite{Tarbert:2007whk} and~\cite{Krusche:2002iq}.
The meaning of the curves and markers is the same as in Fig.~\ref{fig:12C-photo-res}. 
}
\label{fig:40Ca-photo-res}
\end{figure*}

The resemblance between nuclear pion scattering and photoproduction allowed us to employ a unified approach, yielding a satisfactory description of experimental data. 
However, it is important to consider the differences between the two processes when applying the same computational steps.
In the lower panel of Fig.~\ref{fig:scat-photo-corrections-effect}, we show the effects of corrections similar to those illustrated in the upper panel, this time focusing on $\pi^+$-${}^{12}$C elastic scattering in the energy range from threshold to pion laboratory kinetic energy of $\SI{260}{MeV}$.

Figure~\ref{fig:scat-photo-corrections-effect} demonstrates the importance of the final state interaction and modification of the single-nucleon amplitudes in the nuclear medium for both processes.
However, while the scattering amplitude is determined from the Lippmann–Schwinger equation, which is self-consistent, the DWIA photoproduction amplitude, as given by  Eq.~(\ref{FV_gamma-F-def}), is defined as the difference between two terms, each of which in the $\Delta$-resonance energy region is noticeably larger than the total. 
For this reason, the photoproduction amplitude in the $\Delta$-resonance energy region can become sensitive to effects only slightly influencing the scattering process, such as a modification of the $s$-wave part of the pion-nucleus scattering potential.
Consequently, obtaining the correct photoproduction amplitude requires precise and consistent computation of both the photoproduction potential and the scattering amplitude.

As can be seen from Fig.~\ref{fig:scat-photo-corrections-effect}, the energy dependence of the shift in integrated cross sections for photoproduction and scattering, induced by the second-order parts of the potentials $V^{(2)}_\gamma$ and $V^{(2)}$, respectively, displays distinct behaviors.
Although the improvement in the description of the integrated elastic scattering cross section may not be apparent from the lower panel due to scarce data points, significant enhancements become evident when considering the differential elastic scattering cross section~\cite{Tsaran:2023qjx}.
The second-order photoproduction potential $V^{(2)}_\gamma$ consistently increases the cross section across the energy range. 
In contrast, the influence of $V^{(2)}$ exhibits a different behavior, providing a larger effect at lower energies and undergoing a change in sign.
This different energy behavior is explained by the distinct spin structure of the pion scattering and photoproduction amplitudes for individual nucleons, Eqs.~(\ref{CGLN-def}) and~(\ref{t-isospin-struct}).
While the isospin structure of the elementary amplitudes, and consequently the impact on observables from intermediate charge exchange, remains similar for both processes, the photon vertex introduces a larger number of channels involving intermediate nucleon spin-flip.

In Figures~\ref{fig:12C-photo-res} and~\ref{fig:40Ca-photo-res}, we demonstrate a comparison of the integrated and differential cross sections for the coherent $\pi^0$ photoproduction on ${}^{12}$C and ${}^{40}$Ca, respectively, with the experimental data from Refs.~\cite{Tarbert:2007whk, Krusche:2002iq, Gothe:1995zb}.   
The measurements in Ref.~\cite{Tarbert:2007whk} were performed starting from near-threshold photon laboratory energy $\omega^\gamma_\text{lab} = \SI{140}{MeV}$. 
However, for the present analysis, we omit the consideration of the differential cross-section data corresponding to emitted pion laboratory energies below $\SI{30}{MeV}$, as their comprehensive analysis requires the inclusion of nuclear excitations in the propagator $\hat G$ of Eq.~(\ref{Ugamma2nd-def}).
To examine the impact of the second-order corrections, we provide both the predictions of the complete model and the results obtained after setting $V^{(2)}_\gamma = 0$.
The solid red and dashed blue curves in Fig.~\ref{fig:12C-photo-res} match the corresponding curves in Fig.~\ref{fig:scat-photo-corrections-effect}.
For comparison, we also display the theoretical predictions given in Refs.~\cite{Tarbert:2007whk} and~\cite{Krusche:2002iq}, which are based on the model of Ref.~\cite{Drechsel:1999vh}.
In Ref.~\cite{Drechsel:1999vh}, only the first-order photoproduction potential was taken into account, and the effective $\Delta$ self-energy was fitted to experimental data for ${}^4$He~\cite{Rambo:1999jz}.
Due to the mentioned sensitivity of the theoretical results to the scattering amplitude, for the current computation, we have improved the $s$-wave part of the scattering potential for ${}^{40}$Ca (see Appendix~\ref{sec:s-wave-40Ca} for details).

The observed reasonable agreement between the predictions of our model and the experimental data demonstrated in Figs.~\ref{fig:scat-photo-corrections-effect}-\ref{fig:40Ca-photo-res} supports the universality of both our approach, initially applied to pion-nucleus scattering, and the parameter $\Sigma_\Delta$ derived from fitting to the scattering data.
As noted in Ref.~\cite{Krusche:2002iq}, the TAPS data at larger angles possess contamination from incoherent excitations of nuclear levels, which explains the observed disagreement at 200, 290, and $\SI{350}{MeV}$ photon laboratory energy between the data and all theoretical predictions.
The overall discrepancy at $\SI{350}{MeV}$ can be at least partially attributed to the overestimated background contribution in MAID98 at the higher-energy side of the $\Delta$ resonance, as discussed in Sec.~\ref{sec:pion-nucl-photo-ampl}.
As seen from Figs.~\ref{fig:scat-photo-corrections-effect}-\ref{fig:40Ca-photo-res}, there is a disagreement between the data of Refs.~\cite{Tarbert:2007whk} and~\cite{Krusche:2002iq} within the energy range of 200-$\SI{240}{MeV}$ photon laboratory energy for both ${}^{12}$C and ${}^{40}$Ca.
In this energy region, our model better agrees with the more recent Crystal Ball measurements~\cite{Tarbert:2007whk}.

The data for ${}^{16}$O from Ref.~\cite{Tarbert:2007whk} are not taken for the current comparison due to their significant discrepancy with the prior measurements conducted by the Glasgow group using TAPS~\cite{Sanderson:2002tta} and the theoretical predictions based on Ref.~\cite{Drechsel:1999vh}.
This disagreement may stem from the fact that the measurements on ${}^{16}$O were done using a liquid water target, requiring precise treatment for background from hydrogen, which results in additional systematic uncertainties. 
However, our predictions for ${}^{16}$O are found to be in quantitative agreement with the data from Ref.~\cite{Sanderson:2002tta} and the model of Ref.~\cite{Drechsel:1999vh}

\section{Conclusion and outlook}
\label{sec:conclusion}

In the present work, we have developed the second-order momentum space potential for nuclear pion photoproduction on spin-isospin-zero nuclei.
Our approach to photoproduction builds upon our established model for pion-nucleus scattering~\cite{Tsaran:2023qjx}.  
The incorporation of many-body medium effects is achieved through the utilization of the complex effective $\Delta$ self-energy $\Sigma_\Delta$, previously determined by fitting $\pi^\pm$-${}^{12}$C scattering data within the energy range of 80 to \SI{180}{MeV} of pion laboratory kinetic energy.
Since the model parameters originate from the scattering process, the achieved agreement with the experimental data for coherent $\pi^0$ photoproduction on ${}^{12}$C and ${}^{40}$Ca, obtained without any adjustments of $\Sigma_\Delta$, underscores the universality and predictive power of our approach.

The rescattering, involving intermediate excited nuclear states and encompassing intermediate nucleon spin-flip and charge exchange, was incorporated by the second-order part of the photoproduction potential.
This correction yields a non-negligible contribution across the considered energy range, proving essential for describing the experimental data without requiring the additional fitting of the model parameters.
The associated upward shift of the predicted cross section for the coherent $\pi^0$ photoproduction was found to be about 10\%.

Within the $\Delta$-resonance energy region,  the accuracy of the obtained results relies on the precise determination of the pion-nucleus scattering amplitude and is sensitive to the model for the pion photoproduction on a free nucleon.
Moreover, the inclusion of higher-order effects on the effective total $\Delta$ self-energy for both pion scattering and photoproduction may lead to minor discrepancies between both processes.
In principle, one can fit the total $\Delta$ self-energy $\overline \Sigma_\Delta$ for photoproduction to achieve a better agreement with the data.
However, more high-quality data for both scattering and photoproduction are required for such a more refined analysis.

In future work, our intention is to extend the model to heavy nuclei with nonzero isospin for both scattering and photoproduction processes.
The achieved agreement with the photoproduction data also indicates the potential applicability of our demonstrated approach to neutrino-induced pion production on nuclei.

\section*{Acknowledgments}

We express our gratitude to D. Watts and the late B. Krusche for their insightful discussions regarding the experimental data.
We thank F. Colomer and P. Capel for helpful discussions.
This work was supported by the Deutsche Forschungsgemeinschaft (DFG, German Research Foundation), in part through the Collaborative Research Center [The Low-Energy Frontier of the Standard Model, Projektnummer 204404729 - SFB 1044], and in part through the Cluster of Excellence [Precision Physics, Fundamental Interactions, and Structure of Matter] (PRISMA$^+$ EXC 2118/1) within the German Excellence Strategy (Project ID 39083149).

\appendix

\section{In-medium modification of $E_{1+}^{3/2}$}
\label{sec:E1p-in-medium}

\begin{figure}[!thb]
\center{\includegraphics[width=0.99\linewidth]{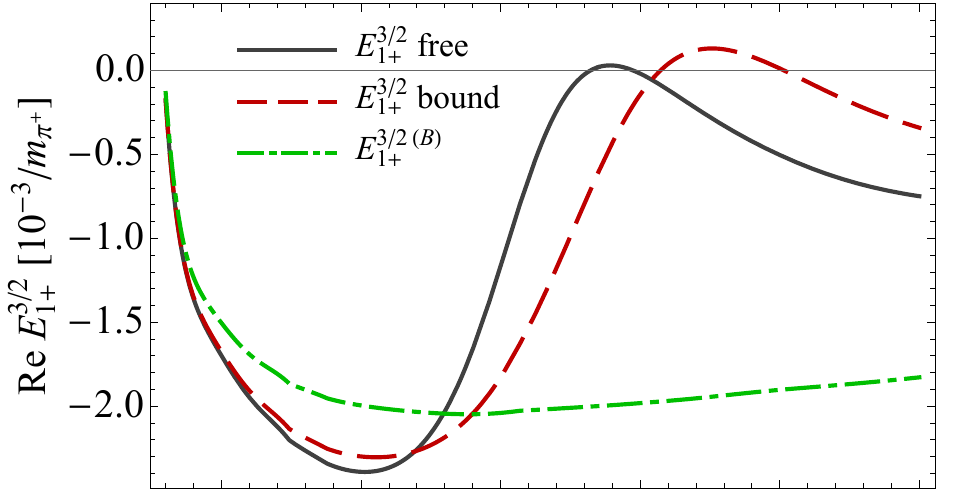}
\includegraphics[width=0.99\linewidth]{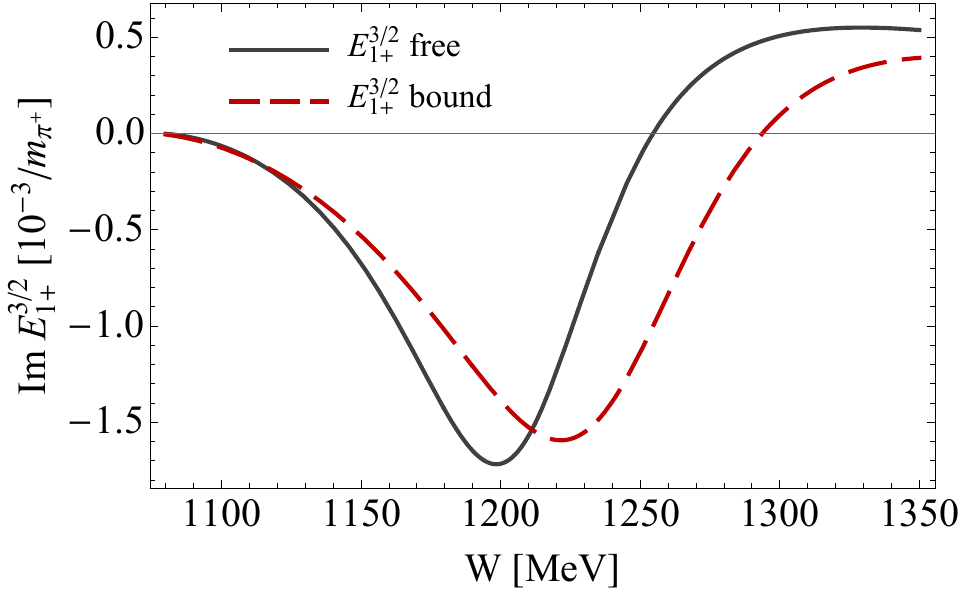}} 
\caption{
The real and imaginary parts of the $E_{1+}^{3/2}$
multipole. 
The meaning of the curves and markers is the same as in Fig.~\ref{fig:M1p3-free-bound}.
}
\label{fig:E1p3-free-bound}
\end{figure}

{

The formalism outlined in Sections~\ref{sec:M1p3-in-medium} and~\ref{sec:Delta-self-energy} can also be directly applied to the $\Delta(1232)$ electric quadrupole $E_{1+}^{3/2}$, which is numerically in the percent range as compared to the dominant magnetic dipole.
By utilizing parameter $\bar E_{3/2} \approx \SI{-0.017}{GeV^{-1/2}}$ (in place of $\bar M_{3/2}$) and setting $n=1$ in Eqs.~(\ref{M1p3-MAID98}) and~(\ref{MAID-f_gammaN}), respectively, we reproduce the free-space resonant part of $E_{1+}^{3/2}$ as given by the MAID98 model.
Employing identical procedures as those for $M_{1+}^{3/2}$, we obtain the in-medium modification for $E_{1+}^{3/2}$.
Finally, in Fig.~\ref{fig:E1p3-free-bound}, we demonstrate the resulting $E_{1+}^{3/2}$ in the nuclear medium along with its free-space counterpart.

Despite the significant change in the amplitude depicted in the plot, its in-medium modification has a negligible effect on the observables. 
For coherent photoproduction, the contribution of $E_{1+}^{3/2}$ is confined to a second-order correction, Eq.~(\ref{V2_gamma-final}), resulting in a shift of less than 1\% in the cross sections.
For the same reason, the in-medium modification of $E_{0+}^\pm$ can also be neglected, which we have also verified by computations.

}

\section{The $s$-wave part of the pion-\texorpdfstring{\ce{^{40}Ca}}{Lg} scattering potential}
\label{sec:s-wave-40Ca}

\begin{figure*}[!th]
\includegraphics[width=0.48\textwidth]{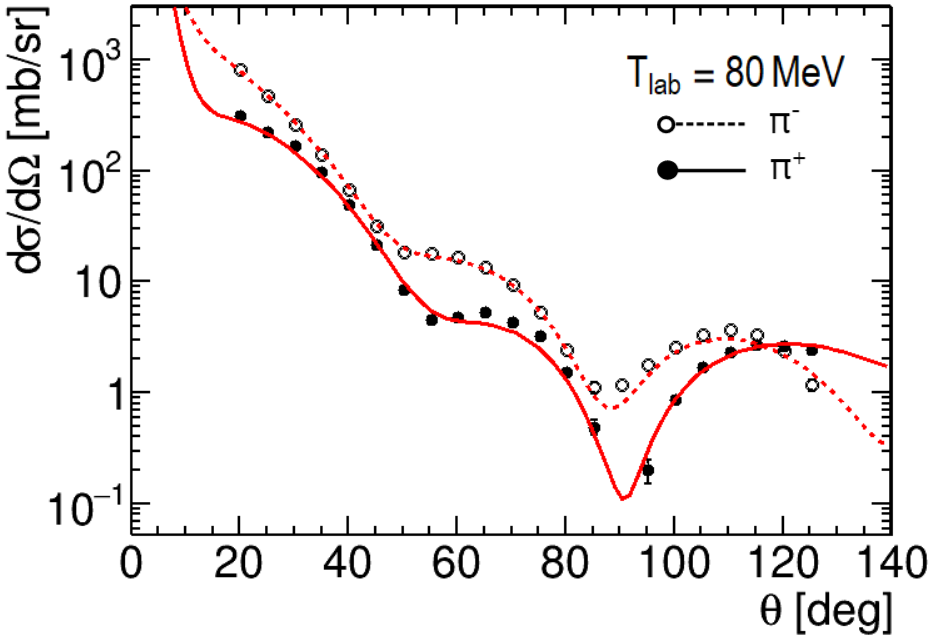}
\includegraphics[width=0.48\textwidth]{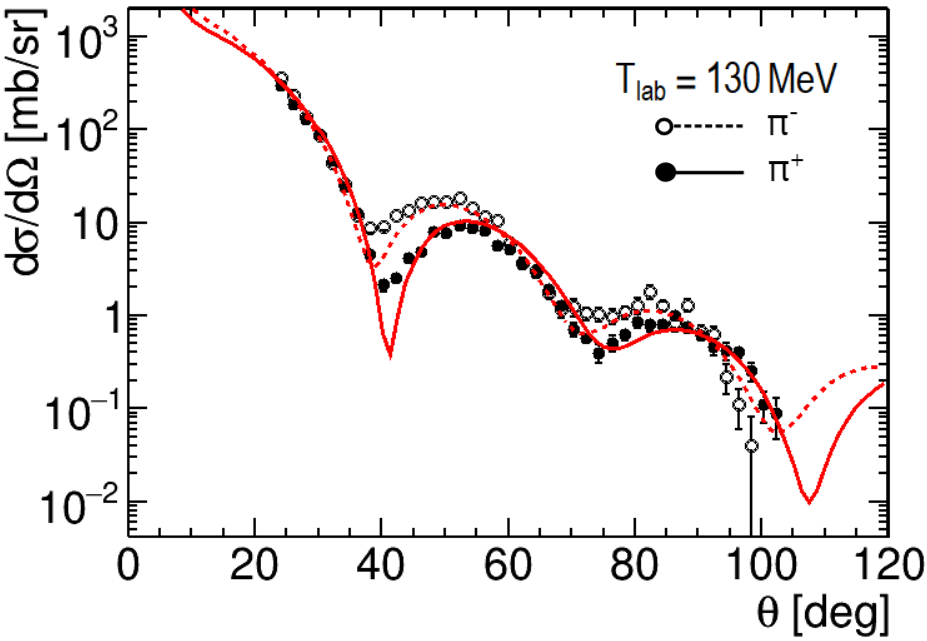}
\includegraphics[width=0.48\textwidth]{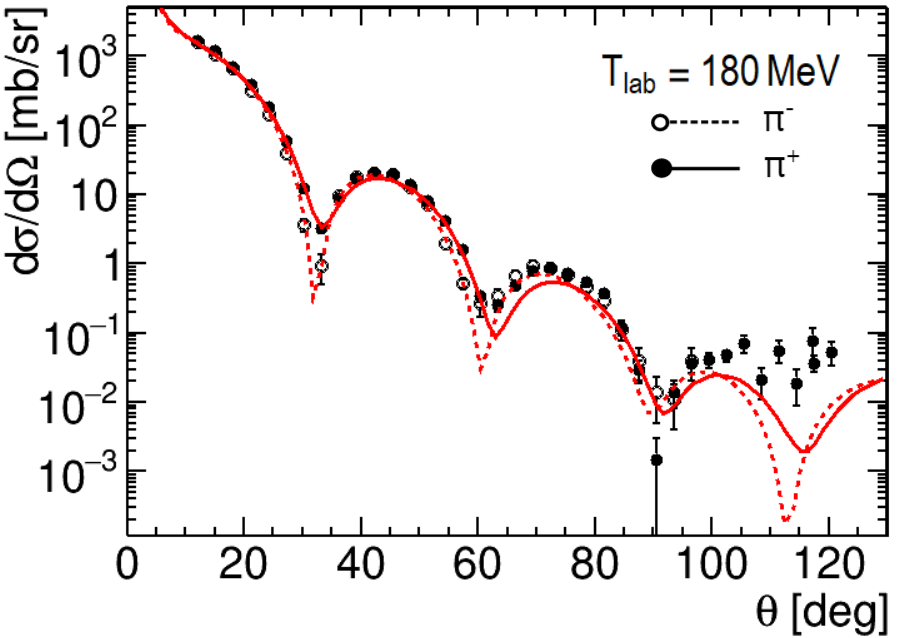}
\caption{
The differential cross section for $\pi^\pm$-${}^{40}$Ca elastic scattering.
Solid curves and closed markers stand for $\pi^+$; dashed and open markers for $\pi^-$.
The experimental data are from Refs.~\cite{Leitch:1984ym} ($\SI{80}{MeV}$) and~\cite{Gretillat:1981bq} ($130, \SI{180}{MeV}$).
}
\label{plt:40Ca_upd_s}
\end{figure*}

In Ref.~\cite{Tsaran:2023qjx}, our research focused on investigating the properties of the $\Delta$ resonance in pion-nucleus scattering.
For this purpose, the $\pi^\pm$-${}^{12}$C scattering data were fitted in the energy range of 80–$\SI{180}{MeV}$ pion laboratory kinetic energy, where the influence of the $\Delta$-resonance excitation is most pronounced.
Within this energy range, theoretical predictions for observables exhibit low sensitivity to variations of the $s$-wave component of the scattering potential.
For this reason, the in-medium modification of the $s$-wave pion-nucleon scattering amplitude was treated within the simplest approximation.
This approach, utilizing model parameters fitted to the data for $\pi^\pm$-${}^{12}$C scattering, also provided a satisfactory description of the differential cross sections for $\pi^\pm$-${}^{40}$Ca elastic scattering.

In this work, however, we aim to further improve upon the accuracy and reliability of the pion-${}^{40}$Ca scattering amplitude.
This is motivated by the well-established sensitivity of the photoproduction process within the $\Delta$-resonance energy region to the nuclear scattering amplitude.
To achieve this improvement, we slightly modify the $s$-wave part of the potential and fit it to the measured differential cross sections for $\pi^\pm$-${}^{40}$Ca elastic scattering, as outlined next.

The $s$-wave part of the momentum space first-order potential for pion scattering on isospin-zero nuclei takes the simple form
\be
V^{(1)}_s(\bm k', \bm k) = \mathscr{W}(\bm k', \bm k)
b_0(k_0)  \rho(\bm q) {v(k') v(k)}.
\label{U1st-fin}
\ee
Here, the complex isoscalar $s$-wave scattering parameter in the nuclear medium is given by
\be
b_0(k_0) = b_0^\text{free}(k_0) + \Delta b_0(k_0),
\ee
where $\Delta b_0$ encapsulates the in-medium modification of the free-space amplitude $b_0^\text{free}$ derived from the SAID analysis~\cite{Workman:2012hx}.
While introducing $\Delta b_0$ also affects both the $s$-$s$- and $s$-$p$-wave interference parts of the second-order pion-nucleus scattering potential ($V_{ss}$ and $V_{sp}$, respectively, in Appendix~C of Ref.~\cite{Tsaran:2023qjx}), this modification yields a negligibly small correction.

In Ref.~\cite{Tsaran:2023qjx}, $\Delta b_0$ was assumed to be purely imaginary and have only one free parameter: the effective $s$-wave isoscalar slope $\alpha$.
In the present calculation for ${}^{40}$Ca, we minimally extend this approach by assuming
\be
\Delta b_0(k) = \beta + i \left(\im\Delta b_0(0) + \alpha \, k\right),
\ee
with $\im\Delta b_0(0) = \SI{0.017}{fm}$ corresponding to the $s$-wave true pion absorption parameter, $B_0$, extracted from the pionic atom analysis of Ref.~\cite{piAF:2022gvw}.
The parameters $\alpha$ and $\beta$ are fitted to the $\pi^\pm$-${}^{40}$Ca scattering data from Refs.~\cite{Leitch:1984ym, Gretillat:1981bq} within the 80-$\SI{180}{MeV}$ energy range .
The resulting values of the parameters are $\alpha = \SI{0.004 \pm 0.009}{fm^2}$ and $\beta = \SI{-0.048 \pm 0.004}{fm}$.

In Fig.~\ref{plt:40Ca_upd_s}, we compare the data with the obtained theoretical fitted curves. 
As a result of the fit, we obtained better agreement at $\SI{80}{MeV}$ pion laboratory kinetic energy, while the plots for 130 and $\SI{180}{MeV}$ show almost identical behavior to our previous results.
This reconfirms the low sensitivity to the $s$-wave part of the potential within the considered energy range. 
For the same reason, introducing additional $s$-wave parameters does not improve our model agreement with the data within the $\Delta$-resonance energy region.

\bibliographystyle{apsrevM}
\bibliography{Bibliography}

\end{document}